\documentclass[a4paper,11pt]{article}
\usepackage{amsmath}
\usepackage{amssymb}
\usepackage{mathtools}
\usepackage{graphicx}
\usepackage{color}
\usepackage{comment}
\usepackage[square,numbers,comma,sort&compress]{natbib}
\usepackage[colorlinks=true,citecolor=blue,linkcolor=blue,urlcolor=blue]{hyperref}
\usepackage{geometry}
\usepackage[normalem]{ulem}
\def\varabstract{ }
\def\varkeywords{ }
\def\vararxivnumber{ }
\def\vartitle{ }
\def\varpreprint{ }
\renewcommand{\title}[1]{\gdef\vartitle{#1}}
\renewcommand{\abstract}[1]{\gdef\varabstract{#1}}
\newcommand{\keywords}[1]{\gdef\varkeywords{#1}}
\newcommand{\arxivnumber}[1]{\gdef\vararxivnumber{#1}}
\newcommand{\preprint}[1]{\gdef\varpreprint{#1}}
\newtoks\authtoks
\renewcommand{\author}[2][]{%
	\authtoks=\expandafter{\the\authtoks#2$^{#1}$\ }%
}
\newtoks\affiltoks
\newcommand{\affiliation}[2][]{%
    \affiltoks=\expandafter{\the\affiltoks{\item[$^{#1}$]#2}}%
}
\newtoks\emailtoks\newcounter{emailcounter}%
\newcommand{\emailAdd}[1]{%
\ifnum\theemailcounter>0\emailtoks=\expandafter{\the\emailtoks, \typeemail{#1}}%
\else\emailtoks=\expandafter{\typeemail{#1}}%
\fi
\stepcounter{emailcounter}}
\newcommand{\typeemail}[1]{\href{mailto:#1}{\tt #1}}

\renewcommand\maketitle{
	\newgeometry{margin=2cm}
	\pagestyle{empty}\setcounter{page}{0}
	\if!\varpreprint!\else\begin{flushright}\varpreprint\end{flushright}\fi
	{\huge\flushleft\sffamily\bfseries\vartitle\par}
\vskip6ex
{\large\bfseries\raggedright\sffamily\the\authtoks\par}
\vskip2ex
\begin{list}{}{%
\setlength{\leftmargin}{0.28cm}%
\setlength{\labelsep}{0pt}%
\setlength{\itemsep}{-3pt}%
\setlength{\topsep}{-\parskip}}
\itshape\small%
\the\affiltoks
\end{list}
\vskip2ex
\noindent\hspace{0.28cm}\begin{minipage}[l]{.9\textwidth}
\begin{flushleft}
\textit{E-mail:} \the\emailtoks
\end{flushleft}
\end{minipage}
\vskip5ex
\noindent{\renewcommand\baselinestretch{.9}\textsc{Abstract:}}\ \varabstract
\vskip5ex 
\if!\varkeywords!\else\noindent{\textsc{Keywords:}}\ \varkeywords \vskip2ex\fi
\if!\vararxivnumber!\else\noindent{\textsc{ArXiv ePrint:}} \href{http://arxiv.org/abs/\vararxivnumber}{\vararxivnumber}\vskip2ex\fi
\newpage
\restoregeometry
\pagestyle{plain}
\hrule
\bigskip\bigskip

{
	\hypersetup{linkcolor=black}
	\tableofcontents
}
\bigskip\medskip
\hrule
\bigskip\bigskip
\setcounter{footnote}{0}
} 

\usepackage[utf8]{inputenc}
\usepackage{verbatim}
\usepackage{textcomp}
\usepackage{nicefrac}
\usepackage{subcaption}
\usepackage{diagbox}
\usepackage{booktabs}
\usepackage{amsmath}
\usepackage{array}
\usepackage{multirow}
\usepackage{slashed}

\usepackage[dvipsnames]{xcolor}
\usepackage{amsmath}
\usepackage{pifont}
   \newcommand{\cmark}{\ding{51}}
   \newcommand{\xmark}{\ding{55}}
\usepackage{booktabs}
\usepackage{amsmath,amssymb}

\global\long\def\mev{\mathrm{MeV}}
\global\long\def\gev{\mathrm{GeV}}
\global\long\def\tev{\mathrm{TeV}}

\usepackage{pdflscape}

\usepackage{subcaption}
\usepackage{verbatim}
\usepackage{textcomp}
\usepackage{nicefrac}
\usepackage{diagbox}
\usepackage{cancel}
\usepackage[normalem]{ulem}
\usepackage{textgreek}
\usepackage{lmodern}
\usepackage{xparse}
\usepackage{booktabs}
\usepackage{caption}
\usepackage{tikz}
\usepackage[compat=1.1.0]{tikz-feynman}
\usepackage{subcaption}
\usepackage{caption}
\usepackage{tikz}
\usepackage[compat=1.1.0]{tikz-feynman}
\usetikzlibrary{calc}
\usetikzlibrary{shapes.misc}
\tikzset{
	>=latex,
    photon/.style={decorate, decoration={snake}, draw=black, thick},
    fermionnoarrow/.style={draw=black, postaction={decorate}, thick},
    scalar/.style={draw=black, postaction={decorate}, decoration={markings,mark=at position .55 with {\arrow{>}}}, thick, dashed},
    scalarnoarrow/.style={draw=black, postaction={decorate},  thick, dashed},
    fermion/.style={draw=black, postaction={decorate},decoration={markings,mark=at position .55 with {\arrow{>}}}, thick},
    gluon/.style={decorate, draw=black, decoration={coil,amplitude=4pt, segment length=5pt}, thick},
    vertex/.style={draw,shape=circle,fill=black,minimum size=3pt,inner sep=0pt},
    fillvertex/.style={draw,shape=circle,fill=black,minimum size=5pt,inner sep=0pt},
    openvertex/.style={draw,shape=circle,minimum size=5pt,inner sep=0pt},
    blob/.style={draw=black,shape=circle,fill=black,minimum size=6pt,inner sep=0pt},
    redvertex/.style={draw=red,shape=circle,fill=red,minimum size=3pt,inner sep=0pt},
    cross/.style={cross out, draw=black,thick, minimum size=5pt, inner sep=0pt, outer sep=0pt}
}

\def\be{\begin{equation}}
\def\ee{\end{equation}}

\DeclareMathOperator{\re}{Re}
\DeclareMathOperator{\im}{Im}

\DeclarePairedDelimiter{\abs}{\lvert}{\rvert}

\makeatletter
\g@addto@macro\bfseries{\boldmath}
\makeatother

\allowdisplaybreaks

\title{Sterile neutrino portals to Majorana dark matter: effective operators and UV completions}

\author[a]{Leonardo~Coito,}
\author[a]{Carlos~Faubel,}
\author[a]{Juan~Herrero-Garc\'ia,}
\author[a]{Arcadi~Santamaria,}
\author[a]{Arsenii~Titov}
\affiliation[a]{Departament de F\'isica Te\`orica, Universitat de Val\`encia 
and IFIC, Universitat de Val\`encia--CSIC, \\
Dr.~Moliner 50, E--46100 Burjassot (Val\`encia), Spain}

\emailAdd{leonardo.coito@uv.es}
\emailAdd{carlos.faubel@uv.es}
\emailAdd{juan.herrero@ific.uv.es}
\emailAdd{arcadi.santamaria@uv.es}
\emailAdd{arsenii.titov@ific.uv.es}

\abstract{
Stringent constraints on the interactions of dark matter with the Standard Model suggest that dark matter does not take part in gauge interactions. In this regard, the possibility of communicating between the visible and dark sectors via gauge singlets seems rather natural. We consider a framework where the dark matter talks to the Standard Model through its coupling to sterile neutrinos, which generate active neutrino masses.  We focus on the case of Majorana dark matter, with its relic abundance set by thermal freeze-out through annihilations into sterile neutrinos. We use an effective field theory approach to study the possible sterile neutrino portals to dark matter. We find that both lepton-number-conserving and lepton-number-violating operators are possible, yielding an interesting connection with the Dirac/Majorana character of active neutrinos. In a second step, we open the different operators and outline the possible renormalisable models. We analyse the phenomenology of the most promising ones, including a particular case in which the Majorana mass of the sterile neutrinos is generated radiatively.}

\keywords{Dark Matter, Sterile Neutrinos, Neutrino Masses, Effective Field Theory, Beyond the Standard Model}
\preprint{FTUV-22-0303.8211 \\
IFIC/22-09}
\arxivnumber{2203.01946}
\makeatother

\begin{document}

\maketitle

\section{Introduction} \label{sec:intro}
%
Non-zero neutrino masses and the existence 
of dark matter (DM) in the Universe constitute 
the two most compelling pieces of evidence 
for physics beyond the Standard Model (SM). 
Many extensions of the SM able to account for
non-zero neutrino masses involve SM gauge singlet fermions $N_R$, 
referred to as sterile or right-handed (RH) neutrinos. 
Apart from the Yukawa coupling $y_\nu$ to the SM singlet operator $LH$, 
with $L$ and $H$ being the $SU(2)_L$ lepton and Higgs doublets, 
respectively, $ N_R $ could also interact with a dark sector
containing, in particular, a DM candidate. 
In this case, $ N_R $ would serve as a portal 
between the visible and dark sectors. 
This is the idea of the sterile neutrino portal to DM~\cite{Pospelov:2007mp}.
Due to the stringent limits on Weakly Interacting Massive Particles (WIMPs) from direct detection (DD), indirect detection (ID) and collider searches, it is natural to consider the case in which the dominant coupling of the dark sector to the SM is via sterile neutrinos.

In what follows, we will assume that 
DM is produced through the thermal freeze-out mechanism. 
Then, depending on the mass of sterile neutrino, $m_N$, one can 
identify two distinct regimes.
If $m_N$ is smaller than the mass of DM, $m_\mathrm{DM}$,
the DM relic abundance is set by its annihilations into sterile neutrinos. 
The latter subsequently decay into SM particles leading to 
ID signatures in photon, charged particle and neutrino spectra. 
This is the so-called secluded regime~\cite{Pospelov:2007mp}. 
The phenomenology of this regime has been studied 
in Refs.~\cite{Escudero:2016ksa,Batell:2017rol,Folgado:2018qlv,Bandyopadhyay:2018qcv} 
considering a simple renormalisable model for the dark sector 
containing a fermion and a real/complex scalar, 
both charged under a $Z_2$/$U(1)$ ``dark'' symmetry. 
In the opposite regime, when $m_N > m_\mathrm{DM}$, 
DM annihilation into a pair of sterile neutrinos is kinematically 
forbidden, unlike the annihilation into SM particles, in particular, 
active neutrinos, which proceeds via active-heavy neutrino mixing. 
In this case, this mixing has to be sizeable to provide 
the annihilation cross section required to explain 
the observed relic abundance. 
The same annihilation process could lead to 
ID signatures at neutrino experiments, see \textit{e.g.}~\cite{Beacom:2006tt,Palomares-Ruiz:2007trf,ElAisati:2017ppn,Olivares-DelCampo:2017feq,Arguelles:2019ouk,BasegmezDuPree:2021fpo}.
The phenomenology of this regime has been investigated 
in Refs.~\cite{Bertoni:2014mva,Batell:2017cmf,Blennow:2019fhy} 
assuming that the dark sector comprises 
a fermion and either a scalar or a vector boson. Furthermore, the freeze-in mechanism of DM production in neutrino portal scenarios has been studied in Refs.~\cite{Chianese:2018dsz,Chianese:2019epo,Bian:2018mkl,Bandyopadhyay:2020qpn}. Further examples of studies investigating DM--neutrino connections can be found in Refs.~\cite{Lindner:2010rr,Ahriche:2016acx,Bhattacharya:2018ljs,Pongkitivanichkul:2019cvm}.

In the present work, we will concentrate on the secluded regime, 
assuming that (i) the DM candidate is a Majorana fermion $\chi$ 
charged under a $Z_2$ symmetry responsible for its stability, 
and (ii) DM interactions with $ N_R $ are given by effective 
four-fermion interactions generated at the new physics scale $\Lambda$. 
The effective field theory (EFT) approach 
for interactions of DM with the SM extended with RH neutrinos 
has been developed in Ref.~\cite{Duch:2014xda},%
\footnote{The EFT of the pure SM (with no RH neutrinos) extended 
with a scalar, fermion or vector DM candidate is well studied;
see \textit{e.g.}~Refs.~\cite{DelNobile:2011uf,DeSimone:2013gj} for early works as well as recent Refs.~\cite{Criado:2021trs,Aebischer:2022wnl} and references therein.}
and the four-fermion operators 
we will focus on in this work form part 
of the basis of dimension-six operators derived in Ref.~\cite{Duch:2014xda}.  
As we will see, restricting ourselves to the interaction of DM with 
the lightest of RH neutrinos, there are three different four-fermion operators.
After studying them in detail, we will discuss simple 
UV completions generating these operators at tree level.
Some of the UV models will lead to other 
operators as well, resulting in either DM and sterile neutrino self-interactions, and/or direct interactions of DM with the SM. 
We will identify the most interesting/promising models 
and study their DM phenomenology, making also a connection to the mechanism of neutrino mass generation.
\begin{figure}[t]
\centering
\includegraphics[width=.9\textwidth]{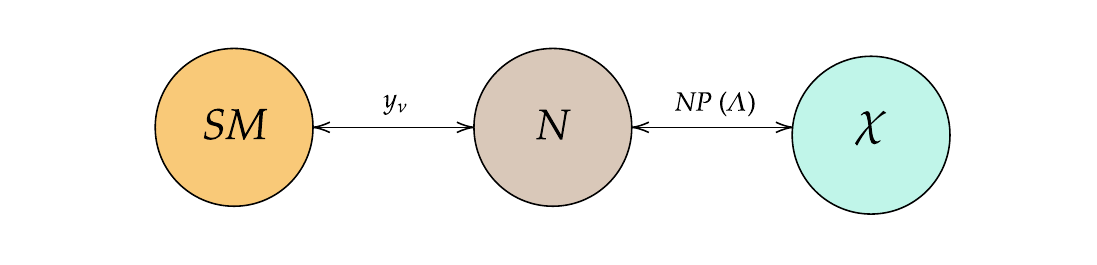}
\caption{Schematic diagram of the considered set-up, in which the sterile neutrino $N$ serves as a portal between the SM and the dark sector containing the DM $\chi$. NP stands for new physics generating 
effective four-fermion interactions between $N$ and $\chi$ at the scale $\Lambda$.}
\label{fig:SM_DM_schematic}
\end{figure}

It is worth noting that what is often meant in the literature 
as neutrino portal is
the operator $(LH)\times\mathcal{O}_\mathrm{dark}$, 
where $\mathcal{O}_\mathrm{dark}$ is 
a singlet of a dark symmetry group~\cite{Falkowski:2009yz}. 
In particular, $\mathcal{O}_\mathrm{dark}$ can be composed of a dark fermion and a dark scalar~\cite{GonzalezMacias:2015rxl,Gonzalez-Macias:2016vxy}. 
The operator $(LH)\times\mathcal{O}_\mathrm{dark}$
may result from integrating out a heavy RH neutrino, 
and thus, the latter is not present in the corresponding EFT.
This is different from what we will discuss 
in the present work. See Fig.~\ref{fig:SM_DM_schematic} for a schematic diagram of our set-up.

The paper is structured as follows. 
In Sec.~\ref{sec:eft}, we describe the EFT approach to the interactions 
of DM with RH neutrinos. 
In Sec.~\ref{sec:open}, we classify the UV models 
generating the four-fermion operators of interest at tree level. 
In Sec.~\ref{sec:phenomenology}, we discuss the phenomenology of several 
selected models. 
In Sec.~\ref{sec:conc}, we summarise our findings and draw our conclusions. 
Finally, some technical details are given in the appendix.

\section{Effective field theory approach}
\label{sec:eft}
%
%
\subsection{Four-fermion sterile neutrino portal operators}  
%
We add to the SM particle content two chiral fermions, $ N_R $ and $\chi_L$, transforming as $(\mathbf{1},\mathbf{1})_0$ 
with respect to the SM gauge group $(SU(3)_C,SU(2)_L)_Y$. The first one, $ N_R $, is the usual RH neutrino, while we consider the case in which the latter, $\chi_L$, has a discrete symmetry $Z_2$ that stabilises it and makes it a potential DM candidate. The most general renormalisable Lagrangian reads
\begin{align} \label{eq:L4}
 \mathcal{L}_4 &= \mathcal{L}_\mathrm{SM} 
 + \overline{ N_R } i \slashed{\partial}  N_R  
 + \overline{\chi_L} i \slashed{\partial} \chi_L - \left[ \frac{1}{2} m_N \overline{ N_R ^c}  N_R  
 + \frac{1}{2} m_\chi \overline{\chi_L} \chi_L^c
 + y_\nu \overline{L} \tilde{H}  N_R 
 + \text{H.c.} \right],
\end{align}
where $\mathcal{L}_\mathrm{SM} $ is the SM Lagrangian (with massless neutrinos). We can always re-phase $ N_R $ and $\chi_L$ in such a way that the masses $m_N$ and $m_\chi$ are real and positive. If the lepton number $U(1)_L$ symmetry is imposed with $L( N_R ) = 1$ and $L(\chi_L) = 0$, then the Majorana mass term for $N$ is forbidden. 
After electroweak symmetry breaking (EWSB), the Higgs field takes a vacuum expectation value (VEV), $\langle H \rangle =\,(0,  v_h/\sqrt{2})^T$, with $v_h=246$ GeV.

In order to generate two non-zero light neutrino masses (either of Dirac or Majorana type), at least two RH neutrinos should be included. For simplicity, in the following  we assume that only one of the sterile neutrinos is lighter than the DM particle, so that $m_N < m_\chi$, and there is an open annihilation channel $\chi \chi \rightarrow  N N$, while the heavier ones decouple from the spectrum. 
Here $\chi$ and $N$ stand for Majorana fields, \textit{i.e.} 
\begin{equation}
\chi=\chi_L+\chi^c_L \qquad 
\text{and} \qquad 
N=N_R^c+N_R\,. 
\label{eq:Majorana_fields}
\end{equation}
We assume that the lightest of the RH neutrinos contributes to active neutrino masses and is also responsible for the DM phenomenology we are interested in. The results can be easily extended to the case of all of them being lighter than DM.

We focus on four-fermion effective operators describing interactions between $\chi$ and $N$, which have masses below the new physics scale $\Lambda$, so that $m_N < m_\chi < \Lambda$. 
At dimension $D=6$, there are three such operators that connect DM with the SM through the sterile neutrinos. 
We will refer to them as \textit{sterile neutrino portal operators} 
or simply \textit{portal operators}.
The corresponding $D=6$ effective Lagrangian reads
\begin{align}
 \mathcal{L}_6 &= \frac{c_1}{\Lambda^2}\, \mathcal{O}_1+ \left[\frac{c_2}{\Lambda^2}\,\mathcal{O}_2+ \frac{c_3}{\Lambda^2}\,\mathcal{O}_3+ \text{H.c.} \right],\label{eq:L_EFT_portalop}
\end{align}
with
\begin{align}
   \mathcal{O}_1 &= (\overline{ N_R } \chi_L)(\overline{\chi_L}  N_R ) 
   = -\frac{1}{2}(\overline{ N_R } \gamma_\mu  N_R )(\overline{\chi_L} \gamma^\mu \chi_L)\,, 
   \label{eq:op1} \\
   \mathcal{O}_2 &=  (\overline{ N_R } \chi_L)(\overline{ N_R } \chi_L) 
   = - \frac{1}{2}(\overline{ N_R }  N_R ^c)(\overline{\chi_L^c} \chi_L)\,, 
   \label{eq:op2} \\
  \mathcal{O}_3 &=(\overline{ N_R ^c}  N_R )(\overline{\chi_L^c} \chi_L) 
  = -\frac{1}{2}(\overline{ N_R ^c} \gamma_\mu \chi_L)(\overline{\chi_L^c} \gamma^\mu  N_R )\,,
  \label{eq:op3}
\end{align}
where in the second equalities we have used Fierz identities.%
\footnote{Note that the r.h.s of Eq.~\eqref{eq:op2} would in general involve 
$+1/2 (\overline{N_R}\sigma_{\mu\nu}N^c_R)(\overline{\chi_L^c}\sigma^{\mu\nu}\chi_L)$. However, such a term vanishes for one generation of $N$ or $\chi$.} 
The Wilson coefficient $c_1$ is real, whereas $c_2$ and $c_3$ are, in general, complex. 

In general, many other operators at mass dimensions $D=5$ and $D=6$ involving $\chi_L$ exist~\cite{Duch:2014xda}. 
In Sec.~\ref{sec:open}, we will consider renormalisable models that generate the portal operators, 
generically together with other ones. 
In Tab.~\ref{tab:operators}, we summarise all the $D \leq 6$ operators generated in these models.
\begin{table}[t]
\centering
\begin{tabular}[t]{lcc}
\toprule
Notation & Operator & Dimension \\
\midrule
\multicolumn{3}{ c }{\textbf{Portal operators}}\\
\midrule
$\mathcal{O}_1$ & 
$(\overline{ N_R } \chi_L)(\overline{\chi_L}  N_R ) $ & 6 \\ 
$\mathcal{O}_2$ & 
$(\overline{ N_R } \chi_L)(\overline{ N_R } \chi_L) $ & 6 \\ 
$\mathcal{O}_3$ & 
$(\overline{ N_R ^c}  N_R )(\overline{\chi_L^c} \chi_L) $ & 6\\
\midrule
\multicolumn{3}{c}{\textbf{Self-interactions}} \\
\midrule
$\mathcal{O}_4$ & 
$(\overline{ N_R ^c}  N_R ) (\overline{ N_R }  N_R ^c) $ & 6 \\ 
$\mathcal{O}_5$ & 
$(\overline{\chi_L^c} \chi_L) (\overline{\chi_L} \chi_L^c)$ & 6 \\
$\mathcal{O}_{\psi \psi}$ & 
$(\overline{\psi} \gamma_\mu \psi) (\overline{\psi} \gamma^\mu \psi)$ & 6\\ 
\bottomrule
\end{tabular}
\qquad
\begin{tabular}[t]{lcc}
\toprule
Notation & Operator & Dimension \\
\midrule
\multicolumn{3}{ c }{\textbf{$N$/DM--SM interactions}} \\
\midrule
$\mathcal{O}_{N H}$ & 
$(\overline{ N_R ^c}  N_R ) (H^\dagger H)$ & 5 \\
$\mathcal{O}_{\chi H}$ & 
$(\overline{\chi_L^c} \chi_L) (H^\dagger H)$ & 5 \\ 
$\mathcal{O}_{N \psi}$ & 
$(\overline{ N_R } \gamma_\mu  N_R ) (\overline{\psi} \gamma^\mu \psi)$ & 6 \\
$\mathcal{O}_{\chi \psi}$ & 
$(\overline{\chi_L} \gamma_\mu \chi_L) (\overline{\psi} \gamma^\mu \psi)$ & 6 \\
\midrule
\multicolumn{3}{c}{\textbf{Majoron interactions}} \\
\midrule
$\mathcal{O}_{\Psi J}$ & 
$(\overline{\Psi} \gamma^\mu \Psi) (\partial_\mu J)$ & 5 \\ 
$\mathcal{O}_{H J}$ & 
$\abs{H}^2 \left(\partial J\right)^2$ & 6 \\ 
\bottomrule
\end{tabular}
\caption{Structure of the effective $D \leq 6$ operators generated by renormalisable models considered in Sec.~\ref{sec:open}. 
All operators, but $\mathcal{O}_{\Psi J}$, are generated by  
integrating out a scalar and/or vector mediator at tree level. 
The operator $\mathcal{O}_{\Psi J}$ 
is a consequence of the non-linear field redefinition defined in Eq.~\eqref{eq:nonlineartransform}.
For this operator, generated in Model B2, $\Psi$ stands for the fields carrying non-zero lepton number, \textit{i.e.}~$\Psi =  N_R $, $\chi_L$, $L$ and $e_R$ (see details in Sec.~\ref{sec:B2}.). 
Finally, in $\mathcal{O}_{\psi\psi},\,\mathcal{O}_{N\psi}$ and $\mathcal{O}_{\chi\psi}$, $\psi$ stands for the SM fermions, \textit{i.e.}~$ 
\psi=L$, $e_R$, $Q$, $u_R$, $d_R$.}
\label{tab:operators}
\end{table}

The operator $\mathcal{O}_1$ preserves lepton number (we will refer to this operator as lepton-number-conserving, LNC). In particular, it is allowed if light neutrinos are Dirac particles. In this case $m_N = 0$, and light neutrino masses are given by $m_\nu = y_\nu v_h/\sqrt{2}$. 
The operators $\mathcal{O}_2$ and $\mathcal{O}_3$ break lepton number in two units (we will refer to these operators as lepton-number-violating, LNV).%
\footnote{As we will see in Sec.~\ref{sec:open}, in some models $\chi_L$ 
is also charged under $U(1)_L$. Depending on its charge, 
$\mathcal{O}_2$ ($\mathcal{O}_3$) will preserve (break) lepton number or vice versa. 
More specifically, if $L(\chi_L) = 1$, then  $\mathcal{O}_2$ is LNC and 
$\mathcal{O}_3$ is LNV, whereas if $L(\chi_L) = -1$, the roles are interchanged. 
In both cases, the LNV operator will break lepton number in four units.
Note that $\mathcal{O}_1$ always preserves lepton number.} 
If lepton number is broken, then in general, also $m_N \neq 0$, and 
in the basis $(\nu_L^c,N_R)$
the neutrino mass matrix reads
\begin{equation}
 \mathcal{M}^\mathrm{tree}_\nu=
 \begin{pmatrix} 
0 & m_{\rm D} \\
m^T_{\rm D} & m_{N}
\end{pmatrix}
\,,
\end{equation}
where $m_{\rm D} = y_\nu v_h/\sqrt{2}$ is the Dirac neutrino mass matrix. 
Assuming for simplicity one generation of active and sterile neutrinos, the eigenvalues of $\mathcal{M}_\nu^\mathrm{tree}$ are
\begin{equation}
m_{1,2}=
\cfrac{1}{2}\left(m_N \pm \sqrt{m_{N}^{2}+4m_{\rm D}^{2}}\right),
\end{equation}
so the active neutrino mass is
\begin{equation}
m_{\nu}=
\cfrac{1}{2}\left|m_{N} - \sqrt{m_{N}^{2}+4m_{\rm D}^{2}}\right|.
\end{equation}
In the following, unless stated otherwise, we assume that there is a contribution to active neutrino masses via the standard seesaw mechanism~\cite{Minkowski:1977sc,Yanagida:1979as,Ramond:1979py,GellMann:1980vs,Glashow:1979nm,Mohapatra:1979ia},%
\footnote{In the context of the SM gauge group, the seesaw mechanism has been discussed in Ref.~\cite{Schechter:1980gr}.}
so that for $m_N\gg m_{\rm D}$, we obtain
\begin{align}
m_{\rm light} &\simeq\,\frac{m_{\rm D}^{2}}{m_{N}}\,, \qquad \nu_{\rm light}  \simeq   \nu_{L}\,, \label{eq:seesaw_rel}\\
m_{\rm heavy}  &\simeq\, m_N\,,\qquad \nu_{\rm heavy} \simeq   N_R \,.
\end{align}
Notice that the mass eigenstates are approximately equal to the weak eigenstates because the neutrino mixing with the heavy states $\sim \sqrt{m_{\rm light}/m_{\rm heavy}}$ is always very small, namely, it is smaller than $10^{-5}$ for $m_N \gtrsim 2$~GeV. This lower limit on $m_N$ stems from the fact that sterile neutrinos should decay before Big Bang nucleosynthesis (BBN) in order not to spoil light-nuclei abundances. 
More specifically, following Ref.~\cite{Batell:2017rol}, we require 
the $N$ lifetime $\tau_N \lesssim 1$~s. 
In the case of Dirac neutrinos (when $N_R$ is the RH component of 
a Dirac neutrino field), the Yukawa coupling $y_\nu \sim 10^{-13}$ is
such that $N_R$ are never in thermal equilibrium with the SM bath 
and their energy density at BBN is negligible, 
leading to no contribution to the number of relativistic degrees of freedom
($N_{\rm eff} \approx 3$), see \textit{e.g.}~Ref.~\cite{Dolgov:2002wy}.

\subsection{Dark matter relic abundance}
%
%
\subsubsection{Thermal equilibrium}
\label{sec:thermal_equi}
%
The only new state that directly couples to the SM is the sterile neutrino, through the Yukawa coupling $y_\nu$. In this scenario, this coupling may be suppressed, as unitarity, EFT validity and DM annihilations require 
\begin{align}
m_N < m_\chi < \Lambda < \mathcal{O}(100)\, \text{TeV}\,,
\end{align}
and therefore, imposing $m_\nu \lesssim 0.05$ eV and using the seesaw relation in Eq.~\eqref{eq:seesaw_rel}, we get
\begin{equation}
 y_\nu \lesssim 1.3 \times 10^{-6} \qquad \text{for} \qquad m_N \lesssim 1~\text{TeV}\,.
\end{equation}
One may wonder whether such a sterile neutrino would be in thermal equilibrium in the early Universe and have standard freeze-out. 
Indeed, such small Yukawa coupling does not suffice to keep $N_R$, 
and thus, DM in equilibrium with the SM all the way down to the freeze-out of DM.%
\footnote{This would not be the case in other variants of the seesaw mechanism, such as inverse~\cite{Mohapatra:1986bd,Bernabeu:1987gr} or linear~\cite{Akhmedov:1995ip,Akhmedov:1995vm} seesaw, where much larger values of the neutrino Yukawa coupling are allowed.}
However,  as we will see in Sec.~\ref{sec:open}, the openings of the effective operators include scalars and gauge bosons that always have interactions with the SM, via Higgs portals and/or kinetic mixing. 
Moreover, other EFT operators, even if irrelevant with respect to DM annihilations compared to those in  Eqs.~\eqref{eq:op1}--\eqref{eq:op3}, may be strong enough to keep DM in thermal equilibrium with the SM. Therefore, it is safe to assume that, early on, $ N_R $ and $\chi_L$ were in thermal equilibrium with the SM.

Even if early on all particles are in equilibrium, 
the dark sector may kinetically decouple from the SM before 
the freeze-out of DM. 
Thermal evolution of a such decoupled dark sector has been studied in Refs.~\cite{Berlin:2016gtr,Binder:2017rgn,Bringmann:2020mgx}.  
The corrections to the cross section needed to obtain the observed relic abundance in this case depend, in particular, on whether 
dark sector particles are relativistic or not at the time of DM chemical freeze-out.
Particularising to our dark sector, if sterile neutrinos are relativistic at the time of kinetic decoupling and down to the freeze-out (\textit{i.e.} for $m_N \lesssim m_\chi/20$), the temperature of the dark sector, 
$T_D$, is similar to that of the SM bath, $T$, leading to a very mild effect on the relic abundance~\cite{Berlin:2016gtr}. 
This is the case realised for the most of the parameter space 
investigated in our analysis in Sec.~\ref{sec:phenomenology}.
On the other hand, if sterile neutrinos and/or dark matter are relativistic at kinetic decoupling but become non-relativistic at freeze-out (\textit{i.e.} for $m_N \gtrsim m_\chi/20$), 
the dark sector will be reheated. In this case, $T_D/T$ could reach a factor of a few~\cite{Berlin:2016gtr}, and the impact on the final relic abundance would be larger. 
In App.~\ref{app:BEqs}, we discuss some implications of 
the departure from the standard assumptions of chemical and kinetic equilibrium.
However, a full treatment of these effects on the relic abundance 
is beyond the scope of the present study.

\subsubsection{Dark matter annihilations}
\label{sec:EFTXS}
%
As there is thermal equilibrium, the DM relic abundance is set by the standard freeze-out of annihilations $\chi \chi \rightarrow N N$. 
In the non-relativistic limit, which is appropriate for freeze-out temperatures, 
the cross section only depends on the relative velocity $v=\left|\vec{v}_1-\vec{v}_2\right|$ of the DM particles 
and can be expanded as:
\begin{equation}
 \sigma v = a + b\frac{v^2}{4} + \mathcal{O}\left(v^4\right).
 \label{eq:Xsec}
\end{equation}
The coefficients $a$ and $b$ are associated to the $s$- and $p$-wave contributions to the annihilations. 
Performing the calculation, we find:
\begin{align}
 a &= \frac{m_\chi^2}{16\pi\Lambda^4} \sqrt{1-r_N^2}\, \left[
 \left(c_1 r_N + 2 \re c_2 + 4 \re c_3\right)^2 
 + 4 \left(\im c_2 - 2 \im c_3\right)^2 \left(1 - r_N^2\right)
 \right],  
 \label{eq:a} \\
 b &= \frac{m_\chi^2}{96\pi\Lambda^4} \frac{1}{\sqrt{1-r_N^2}}\, \Bigg\{
 c_1^2 \left(8 - 28 r_N^2 +23 r_N^4\right) 
 +24 \left[\left(\im c_2 + 2 \im c_3\right)^2 + \left(\re c_2 - 2 \re c_3\right)^2\right] \nonumber \\
  &\phantom{{}={}\frac{m_s^2}{96\pi} \frac{1}{\sqrt{1-r_N^2}}\, \Big\{}
 + 12 r_N^2 \left[(\im c_2)^2+4(\im c_3)^2 - 20\im c_2 \im c_3 - \left(\re c_2 - 6 \re c_3\right) \left(3\re c_2 - 2 \re c_3\right) \right] \nonumber \\
&\phantom{{}={}\frac{m_s^2}{96\pi} \frac{1}{\sqrt{1-r_N^2}}\, \Big\{}
 + 12 c_1 r_N \left(\re c_2 + 2 \re c_3\right) \left(-2 + 3 r_N^2\right) \nonumber \\
 &\phantom{{}={}\frac{m_s^2}{96\pi} \frac{1}{\sqrt{1-r_N^2}}\, \Big\{}
 +12 r_N^4 \left[2\left(\re c_2 - 2 \re c_3\right)^2 - 3 \left(\im c_2 - 2 \im c_3\right)^2\right] \Bigg\}\,,
 \label{eq:b}
\end{align}
with $r_N \equiv m_N/m_\chi$.
In the limit of negligible $m_N$, these formulae reduce to 
\begin{align}
 a &= \frac{m_\chi^2}{4\pi\Lambda^4}\, \left[\abs{c_2}^2 + 4\abs{c_3}^2 + 4 \re (c_2 c_3) \right],
 \label{eq:amR0} \\
  b &= \frac{m_\chi^2}{12\pi\Lambda^4} \Big[c_1^2 + 3 \abs{c_2}^2 + 12 \abs{c_3}^2 - 12\re (c_2 c_3) \Big]\,.
 \label{eq:bmR0}
\end{align}

These results can be understood from the arguments 
based on the discrete symmetries of a pair of Majorana fermions
and conservation of the total angular momentum, $J$. 
They agree with the conclusions of the general analysis performed in Ref.~\cite{Kumar:2013iva}. Below we apply them to our case.

The wave function of the Majorana DM particles in the initial state 
should be anti-symmetric. Since this is defined by  
$(-1)^{L^i}(-1)^{S^i+1} = -1$, 
with $S^i$ and $L^i$ being the spin and the orbital angular momentum 
of the initial pair, we conclude that $L^i+S^i$ must be even.
Choosing the $z$-axis to lie along the direction of motion of the 
outgoing particles, we have $L^f_z =0$ and $J^f_z = S^f_z$, 
where the index $f$ refers to the pair in the final state.
Now, we can study the final states generated by the portal operators in Eqs.~\eqref{eq:op1}--\eqref{eq:op3}. 
\begin{itemize}
 \item If $m_N = 0$, $N_R$ can be described by a Weyl fermion. 
Then, the operator $\mathcal{O}_1$ produces a pair $\overline{ N_R }\,,N_R$, 
with opposite helicities, $+1/2$ for $N_R$ and $-1/2$ for $\overline{N_R}$. Therefore, the spins are aligned and $\abs{S^f_z}=1$.  
Conservation of the total angular momentum implies that $\abs{J^i_z}=\abs{J^f_z}=1$. Since $L^i+S^i$ must be even, the lowest order combination 
that can realise $\abs{J^i_z} = 1$ is $S^i=L^i=1$.
Therefore, we conclude that $\mathcal{O}_1$ generates a $p$-wave suppressed DM annihilation cross section, cf. Eqs.~\eqref{eq:amR0} and \eqref{eq:bmR0}.  
If $m_N\neq0$, there can be a helicity flip and $S^f_z=0$ can be attained 
leading to the $s$-wave proportional to $m_N^2$. 
This agrees with Eq.~\eqref{eq:a}.
Moreover, when the $\overline{ N_R }\,,N_R$ pair is coupled in an $s$-channel, the mediator should be a vector boson. This can also be easily seen by observing the Fierz transformed version of $\mathcal{O}_1$ in Eq.~\eqref{eq:op1}. 
 \item If $m_N=0$, $\mathcal{O}_{2}$ ($\mathcal{O}_{3}$) 
 creates a pair $N_R\,, N_R$ $(\overline{N_R}\,, \overline{N_R})$ 
with both states having the same helicity, $+1/2$ $(-1/2)$, and therefore, 
the spins are anti-aligned and $S^f_z=0$. 
Conservation of $J$ implies that $J^i_z = J^f_z = 0$, 
and at the lowest order we have $S^i = L^i = 0$, 
leading to the $s$-wave DM annihilation cross section. 
This is in agreement with Eq.~\eqref{eq:a}.
Moreover, when the $N_R\,, N_R$ $(\overline{N_R}\,, \overline{N_R})$ pair is coupled in an $s$-channel, the mediator should be a scalar boson. This can also be easily seen by observing the Fierz transformed version of $\mathcal{O}_{2}$ in Eq.~\eqref{eq:op2} and the form of $\mathcal{O}_{3}$ in Eq.~\eqref{eq:op3}. 
\end{itemize}

Furthermore, from Eq.~\eqref{eq:a} we 
observe that the $s$-wave vanishes if $c_1=0$ and $c_2=-2c_3^\ast$.
This can be understood as follows. 
Let us rewrite the Lagrangian in Eq.~\eqref{eq:L_EFT_portalop} in terms of fermion bilinears which have definite transformation properties under
parity, $P$:%
\footnote{Here $\chi$ and $N$ are the Majorana fields defined in Eq.~\eqref{eq:Majorana_fields}.} 
\begin{align}
 \mathcal{L}_{6} = \frac{1}{4\Lambda^2}\Bigg\{&\frac{c_1}{2}\,(\overline{\chi}\gamma_\mu \gamma_5 \chi) (\overline{N}\gamma^\mu \gamma_5 N) +  
 \left(2 \re c_3 - \re c_2 \right) (\overline{\chi}\chi) (\overline{N}N) + 
 \left(2 \im c_3 + \im c_2 \right) (\overline{\chi}\chi) (i \overline{N}\gamma_ 5 N) \nonumber
    \\
   & 
   +\left( \re c_2 + 2 \re c_3 \right) (i \overline{\chi}\gamma_ 5 \chi) (i \overline{N}\gamma_ 5 N) + 
   \left( \im c_2 - 2\im c_3 \right) (i \overline{\chi}\gamma_ 5 \chi) (\overline{N} N) \Bigg\}\,.\label{eq:EFT_discussion_CP}
\end{align}
For a state formed by a pair of Majorana particles, parity is given by
$P=(-1)^{L+1}$, where $L$ 
is the orbital angular momentum 
 of the state. In particular, the bilinear $\overline{\chi}\chi$ annihilates a pair of DM particles with $P=+1$ which at the lowest order corresponds to $L=1$ (see \textit{e.g.}~Ref.~\cite{Kumar:2013iva}),%
\footnote{The same conclusion can be obtained using 
$CP = (-1)^{S+1} = 1$ 
and the antisymmetry of the wave function, which together imply $L=S=1$.} 
hence the DM annihilation cross section $\chi\chi\to NN$ is $p$-wave. Conversely, the bilinear $i\overline{\chi}\gamma_{5}\chi$ annihilates a pair of DM particles with $L=0$, so the DM annihilation cross section is $s$-wave. Finally, the zeroth component of the bilinear $\overline{\chi}\gamma_\mu \gamma_5 \chi$ has $P=-1$, whereas the spatial components have $P=+1$, and thus it contributes to both $s$- and $p$-waves. In view of that, having $p$-wave DM annihilation cross section requires that the terms in the Lagrangian in Eq.~\eqref{eq:EFT_discussion_CP} involving the bilinear $i\overline{\chi}\gamma_{5}\chi$ and $\overline{\chi}\gamma_0 \gamma_5 \chi$ vanish. This implies that $c_1=0$ and $c_2=-2c_3^\ast$. 

For completeness, we also consider the scenario in which 
$ N_R $ is the RH counterpart of the LH SM neutrino $\nu_L$, 
\textit{i.e.} $\nu$ is Dirac with mass $m_\nu \sim 0.05$~eV. 
Lepton number is conserved in this case, and we have only one operator, 
$\mathcal{O}_1$.  The coefficients $a$ and $b$ 
in the annihilation cross section, Eq.~\eqref{eq:Xsec}, are now given by
\begin{align}
 a &= \frac{c_1^2 m_\nu^2}{32\pi\Lambda^4} \sqrt{1 - r_\nu^2}\,,\\
 b &= \frac{c_1^2 m_\chi^2}{192\pi\Lambda^4}\, \frac{\left(16 - 32 r_\nu^2 + 19 r_\nu^4\right)}{\sqrt{1 - r_\nu^2}}\,,
\end{align}
with $r_\nu \equiv m_\nu/m_\chi$. 
Since $m_\nu$ is extremely small,
\begin{equation}
 a \approx 0 \qquad \text{and} \qquad b \approx \frac{c_1^2 m_\chi^2}{12\pi\Lambda^4}\,,
 \label{eq:abDirac}
\end{equation}
to a very good approximation. 
As expected, these expressions agree with those obtained in the Majorana case 
in the limit $m_N \to 0$ with $c_2=0$ and $c_3=0$, cf. Eqs.~\eqref{eq:amR0} and~\eqref{eq:bmR0}.

When the DM annihilation cross section in Eq.~\eqref{eq:Xsec} is thermally averaged one obtains 
an expansion in inverse powers of $x=m_\chi/T$: 
\begin{equation}
 \langle  \sigma v \rangle  = a + \frac{3}{2} b \, x^{-1} + \mathcal{O}\left(x^{-2}\right).
 \label{eq:Xsec_thermal}
\end{equation}
Notice that relativistic corrections will only affect higher order terms in this expansion~\cite{Gondolo:1990dk}.
The typical value for $x$ at freeze-out is around $20$--$25$. 
The observed relic abundance corresponds to 
$\langle  \sigma v \rangle  \approx 2.2 \times 10^{-26}~\text{cm}^3/\text{s}$ 
if $a \neq 0$~\cite{Steigman:2012nb}. 
If the cross section is  
$p$-wave ($a=0$ and $b \neq 0$), 
then $\langle \sigma v \rangle  \approx 4.4 \times 10^{-26}~\text{cm}^3/\text{s}$ is required at freeze-out~\cite{Kolb:1990vq} (see also~\cite{Giacchino:2013bta,Olivares-DelCampo:2017feq}). 
More precisely, the quoted values of $\langle\sigma v\rangle$ 
apply for $m_\chi \gtrsim 10$~GeV, whereas for smaller DM masses 
larger values of $\langle\sigma v\rangle$ are needed~\cite{Steigman:2012nb,Bringmann:2020mgx}.
Hence, in our numerical analysis we employ the results of Ref.~\cite{Bringmann:2020mgx}, 
where $\langle \sigma v \rangle$ that reproduces the observed relic abundance 
is given in Figs.~1 and~4 as a function of $m_\chi$  
for the cases of $s$- and $p$-wave DM annihilation cross section, respectively. 
\begin{figure}[t]
 \centering
 \includegraphics[width=.49\textwidth]{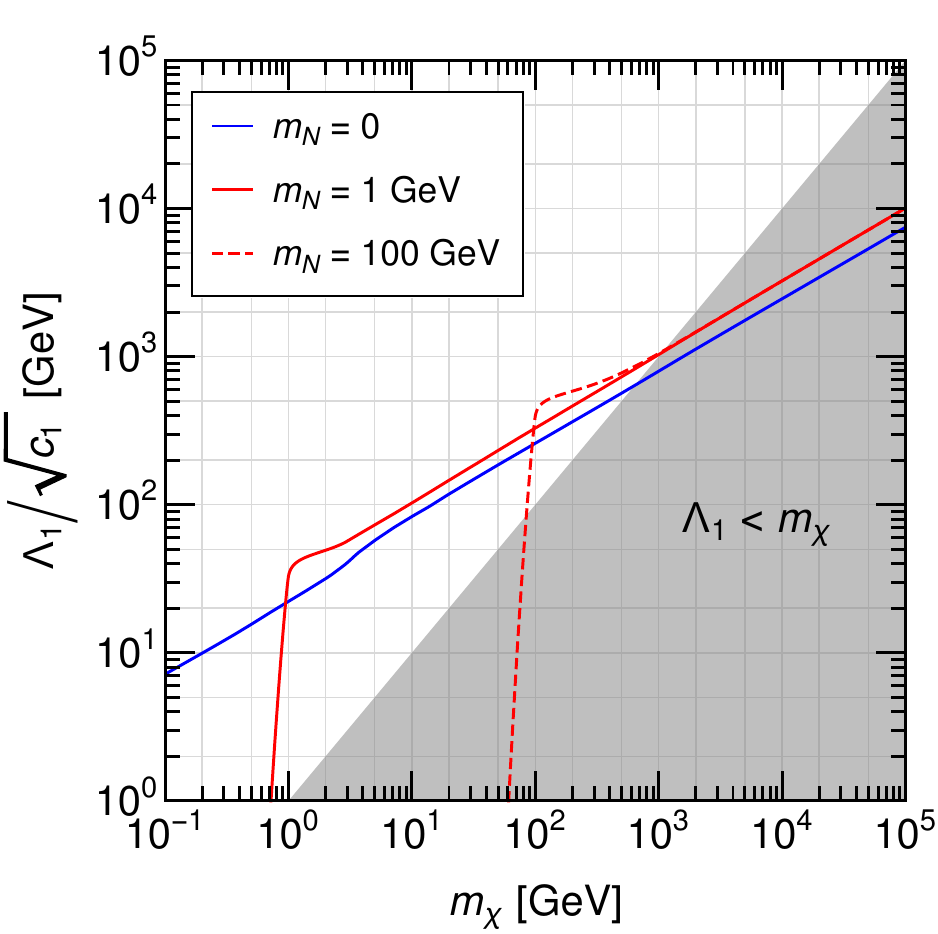}
 \includegraphics[width=.49\textwidth]{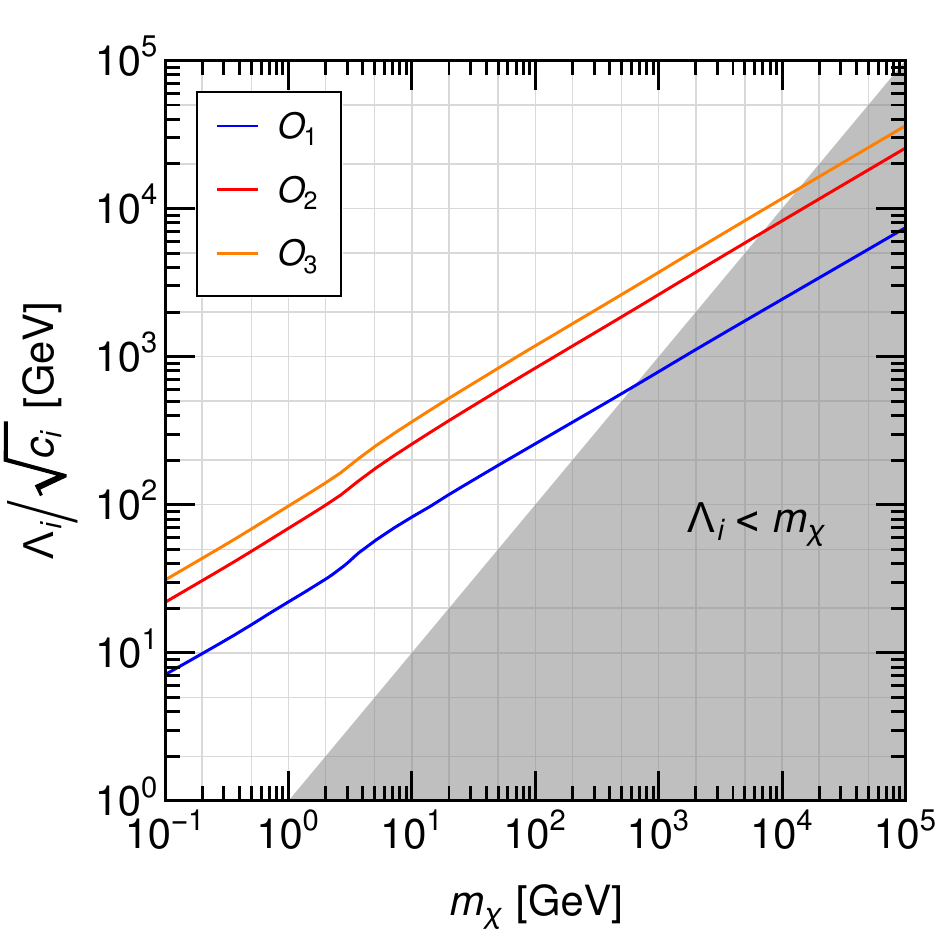}
 \caption{New physics scales needed to obtain the observed relic abundance. Left)~LNC operator, $\mathcal{O}_1$, for different values of the sterile neutrino mass. Right)~All three sterile neutrino portal operators, for $m_N=0$.}
 \label{fig:NPscales}
\end{figure}

In the left panel of Fig.~\ref{fig:NPscales}, 
we show values of $m_\chi$ and the new physics scale $\Lambda_1$ 
(associated with the LNC operator $\mathcal{O}_1$) 
leading to the correct thermal relic cross section assuming standard freeze-out taking place at $x\simeq20$. For $m_N = 0$, the cross section is $p$-wave, 
whereas for $m_N \neq 0$, it is $s$-wave. 
As can be seen, the impact of $m_N$ on the scale of new physics 
needed to get the correct relic abundance is rather moderate. 
For $m_N=0$ (blue line), the lower limit on $m_\chi\simeq 100~\mev$ is set by requiring that $\chi$ is in thermal equilibrium until the freeze-out, which we take at $x\simeq20.$ Near the threshold $m_\chi \simeq m_N$ one can not use the expansion of the DM annihilation cross section in terms of the coefficients $a$ and $b$, as discussed in Ref.~\cite{Gondolo:1990dk}. In this case, we use the general relativistic expression for $\langle \sigma v \rangle$ given in Eq.~(3.8) of the  aforementioned reference (see also~\cite{Claudson:1983js}). We provided this formula in Eq.~\eqref{eq:thermalsigmav_Gondolo} in App.~\ref{app:BEqs}. When we take into account the thermal average for DM masses slightly smaller than $m_N$, the $\langle \sigma v \rangle$ is Boltzmann suppressed. Therefore, we need to decrease $\Lambda_1$ in order to reproduce the observed relic abundance.
In the right panel, we work in the limit $m_N \to 0$ 
and turn on one operator at a time. The LNV operators 
lead to the annihilation cross sections not suppressed by $1/x$, 
cf. Eq.~\eqref{eq:amR0}. 
Thus, for a given DM mass and $\mathcal{O}_2$ ($\mathcal{O}_3$), 
a new physics scale a factor of 3~(4) 
larger than that for $\mathcal{O}_1$ is needed to reproduce 
the observed relic abundance. 
We also display the region where $\Lambda_i < m_\chi$ 
(assuming $c_i = 1$), 
\textit{i.e.}~where the EFT description is not valid.

\section{Tree-level UV completions of the portal operators} 
\label{sec:open}
\begin{figure}[t]
 \centering
 \includegraphics[width=\textwidth]{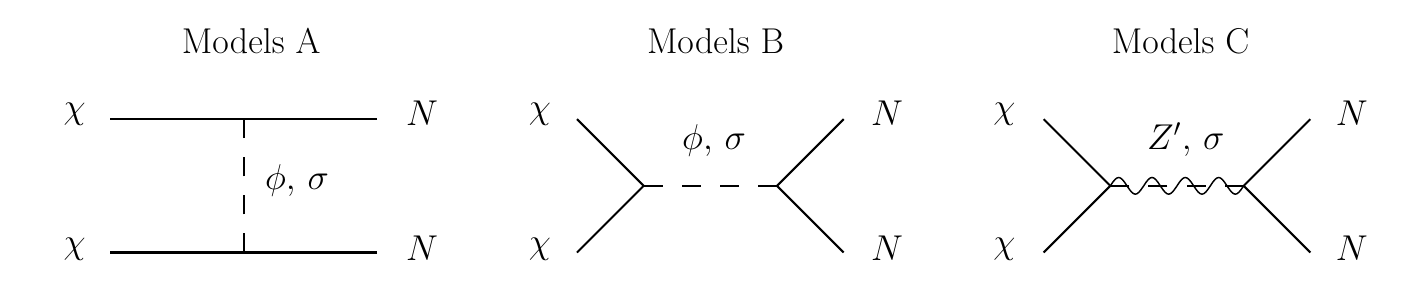}
 \caption{Diagrams for DM annihilation into sterile neutrinos 
 in the three types of renormalisable models considered. 
 $\phi$ and $\sigma$ stand for a real and complex singlet scalar, respectively, whereas $Z^\prime$ is a vector mediator. In Models A, the mediator, exchanged in the $t$-channel, is charged under the $Z_2$. We termed these models \emph{genuine}, as they generate only the four-fermion effective operators with sterile neutrinos and DM.}
 \label{fig:feyn_diagrams}
\end{figure}

%
In this section, we consider tree-level UV completions of the four-fermion neutrino portal operators. They can be divided into models involving a real/complex heavy scalar mediator in either $t$-channel (type A) or $s$-channel (type B), or else a heavy vector mediator (type C), 
as depicted in Fig.~\ref{fig:feyn_diagrams}. In what follows, we assume that the mass of the mediator is always larger than the mass of DM. This forbids annihilation of DM into the mediators and ensures the neutrino portal regime.
In Tab.~\ref{tab:models}, 
we summarise the dark sector particles of each model 
along with their $Z_2$ and $B-L$ charges.
We note that in type-A models, the scalar mediator is charged under the $Z_2$ symmetry stabilising DM, whereas in Models B and C, 
the mediators are neutral under this symmetry. 
In the following, the models that at $D \leq 6$ generate  
only the portal operators $\mathcal{O}_1$ and/or $\mathcal{O}_2$ and/or $\mathcal{O}_3$ will be referred to as \textit{genuine}. 
As we will see below, these are Models A. In addition, 
in Tab.~\ref{tab:classification} 
we provide the tree-level matching conditions for the Wilson coefficients 
of the effective operators generated by the UV models.
See details in the next subsections.
\begin{table}[t]
 \centering
 \begin{tabular}{l  l  c  c}
 \toprule
 Model & Dark sector particles & $Z_2$ & $U(1)_{B-L}$ \\
 \midrule
 \multirow{2}{*}{A1} & Majorana fermion $\chi$ & 
			$1$ & $\phantom{+}0$ \\ 
                               & real scalar $\phi$ & 
			$1$ & $\phantom{+}0$ \\
 \midrule
 \multirow{2}{*}{A2} & Majorana fermion $\chi$ & 
			$1$ & $\phantom{+}0$ \\ 
                               & complex scalar $\sigma$ & 
			$1$ & $-1$ \\
 \midrule
 \multirow{2}{*}{B1} & Majorana fermion $\chi$ & 
			$1$ & $\phantom{+}0$ \\
                               & real scalar $\phi$ & 
			$0$ & $\phantom{+}0$ \\
 \midrule
 \multirow{2}{*}{B2} & chiral fermion $\chi_L$ & 
			$1$ & $+1$ \\
      & complex scalar $\sigma$ & 
			$0$ & $+2$ \\
 \midrule
 \multirow{2}{*}{C1} & Majorana fermion $\chi$ & 
			$1$ & $\phantom{+}0$ \\
       & massive vector boson $Z'$ & 
			$0$ & $\phantom{+}0$\\
 \midrule
 \multirow{3}{*}{C2} & chiral fermion $\chi_L$ & 
			$1$ & $+1$ \\
                                & complex scalar $\sigma$ & 
			$0$ & $+2$ \\
                                & gauge boson $Z'$ & 
			$0$ & $\phantom{+}0$ \\
 \bottomrule
 \end{tabular}
 \caption{Dark sector particles in the UV-complete models together with their 
 $Z_2$ and $B-L$ charges. 
 Under the $Z_2$ symmetry, a field $\varphi \to e^{i\pi k} \varphi$, with $k =0$ or $1$ being the $Z_2$ charge.
In addition to these particles and the SM ones, 
 each model contains RH neutrinos $N_R$, not charged under $Z_2$ and 
 having $B-L$ charge of $-1$. See text for further details.}
 \label{tab:models}
\end{table}

\subsection{Scalar mediator in $t$-channel} 
\label{sec:scalar_med_t_channel}
%
%
\subsubsection{Model A1: real scalar} 
\label{sec:A1}
%
The model includes a real scalar $\phi$, charged under the same $Z_2$ symmetry as $\chi$. The Lagrangian reads
\begin{equation}
 \mathcal{L}_\mathrm{A1} = \mathcal{L}_4
 + \frac{1}{2} \left(\partial_\mu\phi\right) \left(\partial^\mu\phi\right) 
 - V(\phi, H)
 - \left[f \overline{ N_R } \chi_L \phi + \text{H.c.} \right], 
\end{equation}
where $\mathcal{L}_4$ is given in Eq.~\eqref{eq:L4} and $V(\phi, H)$ is the most general scalar potential invariant under $Z_2$:
\begin{equation}
 V(\phi,H) = \frac{1}{2} m_\phi^2 \phi^2 + \lambda_{\phi H} \phi^2 \abs{H}^2 
 + \lambda_\phi \phi^4\,.\label{eq:A1scalarpotential}
\end{equation}
For the sake of simplicity, in what follows, we consider 
one generation of $N_R$ and $\chi_L$, such that the new Yukawa couplings are numbers.
Since we have already re-phased $ N_R $ and $\chi_L$ to render 
the masses $m_N$ and $m_\chi$ real and positive, 
and $\phi$ is a real field, 
the phase of the Yukawa coupling $f$ cannot be absorbed. Hence, in general, it is complex. 
The new Yukawa interaction breaks $U(1)_L$. 
Thus, when integrating out $\phi$, both LNC and LNV operators are
expected to be present. Indeed, we find that $\mathcal{O}_1$ and $\mathcal{O}_2$ are generated with the tree-level matching conditions 
given in Tab.~\ref{tab:classification}. 
No other operators at $D \leq 6$ are induced. 

The DM phenomenology of this model has been studied
in Ref.~\cite{Escudero:2016ksa}; in addition, Ref.~\cite{Tang:2016sib} analyses the cases when $ N $ and $\chi$ have similar masses and Ref.~\cite{Bandyopadhyay:2018qcv} when also $\phi$ does. A wide range of masses for $ N $ and $\chi$ was studied in Ref.~\cite{Batell:2017rol}. On the other hand, the freeze-in mechanism of DM production in this model was considered in Refs.~\cite{Bian:2018mkl,Bandyopadhyay:2020qpn,Chianese:2018dsz}.
\begin{table}[t]
\centering
\begin{tabular}{l|ccc|cc|cc}
\toprule
Model & $c_1/\Lambda^2$ & $c_2/\Lambda^2$ & $c_3/\Lambda^2$ &  
$c_4/\Lambda^2$ & $c_5/\Lambda^2$ & $c_{NH}/\Lambda$ & $c_{\chi H}/\Lambda$ \\
\midrule
A1 & 
$\dfrac{|f|^2}{m_\phi^2}$ & 
$\dfrac{f^2}{2 m_\phi^2}$ & 
\xmark & \xmark & \xmark & \xmark & \xmark \\ 
\midrule
A2a & $\dfrac{f^2}{m_\sigma^2}$ & 
\xmark & \xmark & \xmark & \xmark & \xmark & \xmark \\ 
A2b & $\dfrac{f^2}{m_\sigma^2}$ & 
\xmark & \xmark & \xmark & \xmark & \xmark & \xmark \\ 
A2c & $ 
\dfrac{f^2}{m_\sigma^2}$ & 
$-\dfrac{f^2 \mu_{\sigma}^2}{2m_\sigma^4}$ & 
\xmark & \xmark & \xmark & \xmark & \xmark \\ 
\midrule
B1 & 
\xmark & 
$-\dfrac{2 f^\ast g}{m_\phi^2}$ & 
$\dfrac{f g}{m_\phi^2}$ & 
$\dfrac{\abs{f}^2}{m_\phi^2}$ & 
$\dfrac{\abs{g}^2}{m_\phi^2}$ & 
$\dfrac{f \mu_{\phi H}}{m_\phi^2}$ & 
$\dfrac{g \mu_{\phi H}}{m_\phi^2}$ \\
B2 & 
\xmark & 
$- \dfrac{f g}{m_s^2}$ & 
$\dfrac{fg}{2m_s^2}$ & 
$\dfrac{f^2}{2m_s^2}$ & 
$\dfrac{g^2}{2m_s^2}$ & 
$\dfrac{f \lambda_{\sigma H} v_\sigma}{\sqrt{2} m_s^2}$ & 
$\dfrac{g \lambda_{\sigma H} v_\sigma}{\sqrt{2} m_s^2}$ \\
\midrule
C1 & 
$\dfrac{2 g_N g_\chi}{m_{Z'}^2}$ & 
\xmark & \xmark  & 
 $-\dfrac{g^2_N}{m_{Z'}^2}$ & 
 $-\dfrac{g^2_\chi}{m_{Z'}^2}$ & 
\xmark & \xmark \\
C2 &
$\dfrac{2g'^2 Q_N Q_\chi}{m_{Z'}^2}$ & 
$-\dfrac{f g}{m_s^2}$ & 
$\dfrac{fg}{2m_s^2}$ & 
$\dfrac{f^2}{2m_s^2}-\dfrac{g'^2 Q_N^2}{m_{Z'}^2}$ & 
$\dfrac{g^2}{2m_s^2}-\dfrac{g'^2 Q_\chi^2}{m_{Z'}^2}$ & 
$\dfrac{f \lambda_{\sigma H} v_\sigma}{\sqrt{2} m_s^2}$ & 
$\dfrac{g \lambda_{\sigma H} v_\sigma}{\sqrt{2} m_s^2}$ \\
\bottomrule
\end{tabular}
\caption{Matching conditions for the Wilson coefficients of the effective 
$D \leq 6$ operators generated in renormalisable models 
by integrating out a scalar and/or vector mediator at tree level. 
The structure of the operators is detailed in Tab.~\ref{tab:operators}. 
Notice that in Model B1 with real $g$ and Model B2 
the relation $c_2 = -2c_3^\ast$ is satisfied and annihilations are $p$-wave, 
cf. Eq.~\eqref{eq:a}.
}
\label{tab:classification}
\end{table}
%

\subsubsection{Model A2: complex scalar} 
\label{sec:A2}
%
This model includes a complex scalar, $\sigma=(\rho + i\, \theta)/\sqrt{2}$, charged under both $Z_2$ and $U(1)_L$ (then it is dubbed leptonic scalar). In this case, the Lagrangian reads
\begin{equation}
 \mathcal{L}_\mathrm{A2} = \mathcal{L}_4 
 + \left(\partial_\mu\sigma\right)^\ast \left(\partial^\mu\sigma\right) 
 - V(\sigma, H)
 - \left[f \overline{ N_R } \chi_L \sigma + \text{H.c.} \right], 
 \label{eq:LagA2}
\end{equation}
where $V(\sigma, H)$ is the most general scalar potential preserving lepton number: 
\begin{equation}
 V(\sigma,H) = m_\sigma^2 \abs{\sigma}^2 + \lambda_{\sigma H} \abs{\sigma}^2 \abs{H}^2 + \lambda_\sigma \abs{\sigma}^4\,.
 \label{eq:complex_scalar_potential}
\end{equation}
We can absorb the phase of $f$ in the complex field $\sigma$, 
\textit{i.e.}~$f$ can always be taken real. 
This Yukawa interaction preserves lepton number.
In what follows, we consider three variants of this model 
differing by the status of lepton number: conserved or violated; 
and in the latter case, by the way it is broken.
\begin{itemize}
  \item \textbf{A2a.} Dirac neutrinos, making $m_N=0$ in $\mathcal{L}_4$. 
  Lepton number is conserved, and only $\mathcal{O}_1$ is generated. In this case, annihilations are effectively $p$-wave, see Eq.~\eqref{eq:abDirac}. 
  Therefore, indirect limits are avoided and low DM masses ($\lesssim 10$ GeV) are allowed. Ultimately, the lower limit $m_\chi\simeq 100~\mev$ is set by the DM being in thermal equilibrium until the freeze-out, which we assume at $x\simeq20$, see Fig.~\ref{fig:NPscales}.
   
   This is an interesting scenario, however, it is difficult to test it, as neutrinos are Dirac particles, and DD and ID are suppressed. If no signal is observed 
neither in DM searches nor in neutrinoless double beta decay experiments, 
this would remain a valid option.

  \item \textbf{A2b.} Majorana neutrinos, with $m_N$ being a free parameter. Lepton number is violated by $m_N$, while the interaction Lagrangian (including the potential) is LNC. 
Thus, from the EFT point of view, the operators $\mathcal{O}_2$ and $\mathcal{O}_3$ are not induced. The relic abundance obtained via the freeze-in mechanism for this model was studied in Refs.~\cite{Cosme:2020mck,Coy:2021sse,Becker:2018rve}.

   \item \textbf{A2c.} Majorana neutrinos, but with $m_N= 0$ in Eq.~\eqref{eq:L4} and lepton number being \textit{softly} broken only in the scalar potential:
 \begin{equation}
   \mathcal{L}_{\mathrm{A2 c}} = \mathcal{L}_{\mathrm{A2}} \rvert_{m_N = 0}  - \left[\frac{1}{2}\mu^2_\sigma\, \sigma^2 + \,\text{H.c.}\right]. 
   \label{eq:LagA2c}
\end{equation}
We can absorb
the phase of $\mu_\sigma^2$ in $\sigma$, hence making $\mu_\sigma^2$ 
real and positive. 
Further, in the absence of $m_N$, we can redefine $N_R$ 
(as well as $L$ and $e_R$)
in such a way that $f$ also becomes real.
 This model has some interesting features. For example, 
 finite $m_N$ is generated at one loop. We will discuss this feature
 in Sec.~\ref{sec:mnu_1loop}. 
 Integrating out the complex scalar $\sigma$, we find that both 
 $\mathcal{O}_1$ and $\mathcal{O}_2$ are generated with the matching conditions 
 given in Tab.~\ref{tab:classification}. 
 As expected, the Wilson coefficient of $\mathcal{O}_2$ is proportional 
 to the LNV parameter $\mu_\sigma^2$. 
Alternatively, we can integrate out the real, $\rho$, 
 and imaginary, $\theta$, parts of $\sigma$. This leads to the following matching relations:
 \begin{equation}
  \frac{c_1}{\Lambda^2} = \frac{f^2}{2} 
 \left(\frac{1}{m_\rho^2} + \frac{1}{m_\theta^2}\right) 
 \qquad \text{and} \qquad
 \frac{c_2}{\Lambda^2} = \frac{f^2}{4} 
 \left(\frac{1}{m_\rho^2} - \frac{1}{m_\theta^2}\right)\,,
 \end{equation}
 where
\begin{equation}
 m_\rho^2 = m_\sigma^2 + \mu_\sigma^2 
 \qquad \text{and} \qquad 
 m_\theta^2 = m_\sigma^2 - \mu_\sigma^2\,. 
 \label{eq:masses}
\end{equation}
For $\mu_\sigma^2/m_\sigma^2 \ll 1$, these matching conditions 
reduce to those given in Tab.~\ref{tab:classification}.
 No other non-renormalisable operators of $D \leq 6$ are induced.
\end{itemize}

\subsection{Scalar mediator in $s$-channel}
%
%
\subsubsection{Model B1: real scalar} 
\label{sec:B1}
%
It includes a real scalar $\phi$, this time not charged under the $Z_2$ symmetry. In this case, 
\begin{equation}
 \mathcal{L}_\mathrm{B1} = \mathcal{L}_4 
 + \frac{1}{2} \left(\partial_\mu\phi\right) \left(\partial^\mu\phi\right) 
 - V(\phi, H)
 - \left[f \overline{N_R^c}  N_R \,\phi + g \overline{\chi^c_L} \chi_L\,\phi +  \,\text{H.c.}\right], \label{eq:B1_Lagrangian}
 \end{equation}
 where $V(\phi, H)$ is the most general scalar potential:%
 \footnote{We note that the term linear in $\phi$ can always be removed 
 by a shift.} 
 \begin{equation}
  V(\phi,H) = \frac{1}{2}m^2_\phi\,\phi^2 + \mu_{\phi}\phi^3 + \lambda_\phi\,\phi^4 + \mu_{\phi H}\phi|H|^2 +\lambda_{\phi H}\phi^2|H|^2\,.\label{eq:B1_potential}
 \end{equation}
In general, the Yukawa couplings $f$ and $g$ are complex. 
Integrating out $\phi$ leads to the LNV portal operators $\mathcal{O}_2$ 
and $\mathcal{O}_3$.
As can be inferred from the matching conditions given in Tab.~\ref{tab:classification}, 
in the case of real $g$, the relation $c_2 = -2c_3^\ast$ holds. 
Thus, annihilations are $p$-wave (since $c_1 = 0$), cf. Eq.~\eqref{eq:a}. 
This agrees with the conclusion of the discussion on $P$ 
of the initial pair of Majorana DM particles (see Sec.~\ref{sec:EFTXS}).
 
Apart from $\mathcal{O}_2$ and $\mathcal{O}_3$, we have four more operators at $D \leq 6$ when integrating out $\phi$. 
Namely, there are two $D=5$ operators:
 \begin{align}
  \mathcal{O}_{NH} &= (\overline{ N_R ^c}  N_R ) (H^\dagger H)\,, 
  \label{eq:opnuH} \\
  \mathcal{O}_{\chi H} &= (\overline{\chi_L^c} \chi_L) (H^\dagger H)\,. 
  \label{eq:opchiH}
 \end{align}
As can be seen from Tab.~\ref{tab:classification}, the Wilson coefficients of these operators are controlled by the scalar coupling 
$\mu_{\phi H}$. Thus, if $\mu_{\phi H}$ is small, these operators are suppressed with respect to the neutrino portal operators.
Upon EWSB, the first operator contributes to the (mostly-)sterile neutrino mass 
and Higgs or $N$ decays, depending on the value of $m_N$. 
The second operator contributes to the DM Majorana mass 
and provides the fermionic Higgs portal with associated
DM phenomenology, see \textit{e.g.}~\cite{Lopez-Honorez:2012tov,Fedderke:2014wda,GAMBIT:2018eea}.  
In addition, at $D=6$, we find the four-fermion self-interactions:
\begin{align}
 \mathcal{O}_4 &= (\overline{ N_R ^c}  N_R ) (\overline{ N_R }  N_R ^c) 
 = \frac{1}{2} (\overline{ N_R } \gamma_\mu  N_R ) (\overline{ N_R } \gamma^\mu  N_R )\,,\label{eq:op4} \\
 \mathcal{O}_5 &= (\overline{\chi_L^c} \chi_L) (\overline{\chi_L} \chi_L^c) 
 = \frac{1}{2} (\overline{\chi_L} \gamma_\mu \chi_L) (\overline{\chi_L} \gamma^\mu \chi_L)\,\label{eq:op5},
\end{align}
with the matching conditions provided in Tab.~\ref{tab:classification}. Since we consider one generation of $ N_R $ and $\chi_L$, 
the operators $(\overline{ N_R ^c}  N_R ) (\overline{ N_R ^c}  N_R )$ 
and $(\overline{\chi_L^c} \chi_L) (\overline{\chi_L^c} \chi_L)$ 
vanish identically.

\subsubsection{Model B2: global $U(1)_{B-L}$} 
\label{sec:B2}
%
Instead of a real scalar $\phi$, this model includes a complex scalar $\sigma$. 
A complex scalar calls for an associated $U(1)$ symmetry. 
In this case,  
we will consider lepton number, or rather $U(1)_{B-L}$, 
since the latter is an anomaly-free global symmetry of the SM. 
The corresponding lepton charges are $L(N_R ) = L(\chi_L^c) = 1$~%
\footnote{Other charge assignments are possible since the new fermions are singlets under the SM gauge group.}
and $L(\sigma) = -2$. The Lagrangian of this model is given by
\begin{equation}
 \mathcal{L}_\mathrm{B2} = \mathcal{L}_4 \rvert_{m_N = m_\chi = 0} 
 + \left(\partial_\mu\sigma\right)^\ast \left(\partial^\mu\sigma\right) - V(\sigma, H) 
 -  \left[f \overline{N_R^c}  N_R  \sigma + g \overline{\chi_L} \chi_L^c \sigma 
 +\text{H.c.}\right], 
 \label{eq:LagB2}
\end{equation}
where $V(\sigma,H)$ is the LNC potential 
given in Eq.~\eqref{eq:complex_scalar_potential}. 
The couplings $f$ and $g$ can be rendered real.
This model has been studied in detail in Ref.~\cite{Escudero:2016tzx}.
If the complex scalar acquires a VEV, $v_\sigma$, 
the $U(1)_{B-L}$ symmetry gets broken spontaneously, 
and Majorana masses
$m_N = \sqrt{2} f v_\sigma$ and $m_\chi = \sqrt{2} g v_\sigma$ are generated. 

We can parameterise the complex scalar as
\begin{equation}
 \sigma = \frac{1}{\sqrt{2}}\left(v_\sigma + s\right) e^{iJ/v_\sigma}\,.
 \label{eq:sigmaEXP}
\end{equation}
Then $J$ corresponds 
to the (massless) Goldstone boson, the Majoron, and $s$ is the radial excitation. 
In this parameterisation, $J$ is not present in the potential 
and appears in the Lagrangian only through the kinetic term
\begin{equation}
 \left(\partial_\mu\sigma\right)^\ast \left(\partial^\mu\sigma\right) = 
 \frac{1}{2} \left(\partial s\right)^2 + \frac{1}{2} \left(\partial J\right)^2 
 + \frac{1}{v_\sigma} s \left(\partial J\right)^2 + \frac{1}{2v_\sigma^2} s^2 \left(\partial J\right)^2\,,
\end{equation}
and the Yukawa interactions in Eq.~\eqref{eq:LagB2}.
Further, we can rotate the fields carrying non-zero lepton number, 
namely, $\Psi =  N_R $, $\chi_L^c$, $L$ and $e_R$, as
\begin{equation}
 \Psi \to e^{-i J/(2v_\sigma)} \Psi\,,
 \label{eq:nonlineartransform}
\end{equation} 
and remove $J$ from all Yukawa interactions. 
After this transformation, the kinetic terms for the fermions $\Psi$ 
will induce
\begin{equation}
 \mathcal{O}_{\Psi J} = (\overline{\Psi} \gamma^\mu \Psi) (\partial_\mu J) 
 \qquad \text{with} \qquad 
 \frac{c_{\Psi J}}{\Lambda} = \frac{1}{2v_\sigma}\,.\label{eq:B2_D5_dJ}
\end{equation}
In this way, the derivative nature of Goldstone boson's couplings is manifest.
It is worth noting that, despite having $D=5$, this operator is not related 
to integrating out a heavy mediator, but is a consequence of the non-linear 
field redefinition performed in Eq.~\eqref{eq:nonlineartransform}.

In the spirit of the EFT approach we are pursuing in the current study, it is interesting to see which effective operators are generated 
at low energies if $s$ is heavy 
(for concreteness, we assume that its mass is larger than the electroweak scale). In what follows, we work in the unbroken phase of the electroweak symmetry. Minimising the potential in Eq.~\eqref{eq:complex_scalar_potential} leads to $m_\sigma^2 = - \lambda_\sigma v_\sigma^2$. Furthermore, we find the mass of 
the radial excitation to be
$m_s^2 = 2 \lambda_\sigma v_\sigma^2$. 
Integrating out $s$, to order $\mathcal{O}(1/m_s^2)$ 
we find the portal operators $\mathcal{O}_{2,3}$, 
self-interactions $\mathcal{O}_{4,5}$ 
and the operators $\mathcal{O}_{NH, \chi H}$, 
with the matching conditions for their Wilson coefficients 
summarised in Tab.~\ref{tab:classification}. 
Interestingly, $c_2 = -2c_3^\ast$, leading (since $c_1 = 0$) 
to a $p$-wave annihilation cross section for $\chi\chi \to NN$, 
cf. Eq.~\eqref{eq:a}. 
Similarly to Model B1, $c_{NH}$ and $c_{\chi H}$ are controlled by an 
independent scalar coupling, $\lambda_{\sigma H}$, and thus, 
$\mathcal{O}_{NH}$ and $\mathcal{O}_{\chi H}$ are suppressed 
if $\lambda_{\sigma H} \ll 1$.

On top of the interactions given in Tab.~\ref{tab:classification}, 
there is a $D=6$ operator describing the 
Higgs--Majoron interaction (cf. Ref.~\cite{Coito:2021fgo}):
\begin{equation}
 \mathcal{O}_{HJ} = \abs{H}^2 \left(\partial J\right)^2 
 \qquad \text{with} \qquad 
 \frac{c_{HJ}}{\Lambda^2} = - \frac{\lambda_{\sigma H}}{m_s^2} = - \frac{\lambda_{\sigma H}}{2 \lambda_\sigma v_\sigma^2}\,.\label{eq:B2_D6_dJ}
\end{equation}
It is interesting to note that the $\abs{H}^6$ operator is not 
generated at tree level due to a peculiar cancellation 
coming from the $s^3$ and $s^2 \abs{H}^2$ terms in the potential 
upon using the equation of motion for $s$ and the relation between 
$m_s$ and $v_\sigma$. 
(This had been previously noted in Ref.~\cite{Gorbahn:2015gxa}.)
Finally, the parameters of the SM potential, 
\begin{equation}
 V_\mathrm{SM} = m_H^2 \abs{H}^2 + \lambda_H \abs{H}^4\,,
\end{equation}
get shifted as 
\begin{equation}
 m_H^2 \to m_H^2 + \frac{1}{2} \lambda_{\sigma H} v_\sigma^2 
 \qquad \text{and} \qquad 
 \lambda_H \to \lambda_H - \frac{\lambda_{\sigma H}^2}{4 \lambda_\sigma}\,.
\end{equation}
%

\subsection{Vector mediator}
%
%
\subsubsection{Model C1: massive vector boson} 
\label{sec:C1}
%
The vector form of the operator $\mathcal{O}_1$, see Eq.~\eqref{eq:op1}, 
suggests that it can be generated by the exchange of a heavy neutral vector boson, $Z'_\mu$. The Lagrangian that could (effectively) describe such an exchange has the following form:
\begin{equation}
 \mathcal{L}_\mathrm{C1} = \mathcal{L}_4 
 - \frac{1}{4} Z'_{\mu\nu} Z'^{ \mu\nu} + \frac{1}{2} m_{Z'}^2 Z'_\mu Z'^\mu 
 + g_N \overline{ N_R } \gamma^\mu  N_R  Z'_\mu + g_\chi \overline{\chi_L} \gamma^\mu \chi_L Z'_\mu\,, 
 \label{eq:LagC1}
\end{equation}
where $Z'_{\mu\nu}$ is the corresponding field strength tensor. 
In general, kinetic mixing among $Z'$ and $Z$, $\epsilon Z'_{\mu\nu} Z^{\mu\nu}$, 
as well as mass mixing, $\delta m^2 Z'_\mu Z^\mu$, are also allowed.  
However, to ensure the neutrino portal regime (and reduce the number of independent parameters), we will set $\epsilon$ and $\delta m^2$ to zero. 
The couplings $g_N$ and $g_\chi$ are real.
Integrating out $Z'$, we obtain the LNC operator $\mathcal{O}_1$ 
and the four-fermion self-interactions $\mathcal{O}_4$ and $\mathcal{O}_5$  
defined in Eqs.~\eqref{eq:op4} and \eqref{eq:op5}. 
The matching relations for the respective Wilson coefficients are provided in Tab.~\ref{tab:classification}.

The Lagrangian in Eq.~\eqref{eq:LagC1} should be viewed as an effective 
description of the interaction mediated by a massive vector boson.%
\footnote{For the conditions of applicability of such type of models see \textit{e.g.} Ref.~\cite{Kahlhoefer:2015bea}.} 
To have a UV-complete gauge-invariant model, 
$ N_R $ and $\chi_L$ should be charged under the same gauge symmetry, 
namely, $U(1)_{B-L}$. 
This brings us to the next option: a gauged version of Model B2.

\subsubsection{Model C2: gauged $U(1)_{B-L}$} 
\label{sec:C2}
%
It is well known that promoting $U(1)_{B-L}$ to a local symmetry 
requires the addition to the SM particle content of three RH 
neutrinos to cancel gauge anomalies.%
\footnote{This is not the only accidental (global) symmetry of the SM that can be gauged. It is well known that differences of individual lepton flavour numbers, such as $L_\mu-L_\tau$, are also 
anomaly free in the pure SM (with no RH neutrinos)~\cite{He:1990pn}. In SM extensions with additional fermions (like our scenario), $L_\mu-L_\tau$ and its variants can also be gauge symmetries if the new fermions have the proper charge assignments. However, flavour symmetries have strong implications for neutrino masses and mixings, and thus, deserve further studies. Therefore, although they may have interesting implications, in the following we discuss the flavour-blind symmetry $B-L$. 
Let us stress that a distinct feature of this gauge symmetry is that cancellation of gauge anomalies calls for the addition to the SM of 
three chiral fermions, unlike the cases of gauged differences of individual lepton flavour numbers.} 
In the considered case, one of them is traded by the chiral fermion $\chi_L^c$ 
odd under $Z_2$. This is why it is important to have $L( N^{1,2}_R ) = L(\chi_L^c) = 1$. 
For concreteness, we assume that one of the two sterile neutrinos 
is lighter than DM, whereas the second one has a mass around the scale of 
$U(1)_{B-L}$ symmetry breaking, $v_\sigma$.
The Lagrangian of this model reads%
\footnote{In general, the kinetic mixing term among $Z'$ and $B$, $\epsilon Z'_{\mu\nu} B^{\mu\nu}$, 
is allowed. Here $B^{\mu\nu}$ is the field strength tensor of $U(1)_Y$. However, to ensure the neutrino portal regime we assume that the physical kinetic mixing is negligible.}
\begin{equation}
 \mathcal{L}_\mathrm{C2} = \mathcal{L}_\mathrm{B2} - \frac{1}{4} Z'_{\mu\nu} Z'^{ \mu\nu} \,, 
 \label{eq:LagC2}
\end{equation}
where in $\mathcal{L}_\mathrm{B2}$
the (covariant) derivatives are modified to include the piece 
associated with the new gauge symmetry:
\begin{equation}
 D_\mu = D_\mu^\mathrm{SM} - i g' Q_{B-L} Z'_\mu\,,
\end{equation}
with $g'$ and $Q_{B-L}$ being, respectively, 
the new gauge coupling and 
the $B-L$ charge of the field $D_\mu$ acts upon. 
We provide the $B-L$ charges of this model's fields in Tab.~\ref{tab:BLcharges}.
Upon spontaneous breaking of $U(1)_{B-L}$, $Z'$, $ N_R $ and $\chi_L$ acquire 
their masses: $m_{Z'} = 2 g' v_\sigma$, $m_N = \sqrt{2} f v_\sigma$ 
and $m_\chi = \sqrt{2} g v_\sigma$. We assume that $m_N<m_\chi<m_Z^\prime$ which implies that $f<g<\sqrt{2}g^\prime$.  

From an EFT point of view, if $v_\sigma$ is larger than the weak scale 
and the couplings $g'$ and $\lambda_\sigma$ are not too small, 
we can integrate out both $s$ and $Z'$. 
It is convenient to go to the unitary gauge, 
rendering $\sigma$ from Eq.~\eqref{eq:sigmaEXP} real in each point of spacetime. In this gauge, the would-be Goldstone boson $J$ is removed from the theory. We list in Tab.~\ref{tab:classification} the Wilson coefficients 
of the operators generated in the EFT. 
As in Model B2, we find $c_2 = -2 c_3^\ast$, so the contributions 
of these operators to the $s$-wave part of the cross section 
for $\chi\chi \to NN$ cancel, see Eq.~\eqref{eq:a}. 
It is interesting to note that $c_4$ and $c_5$ vanish if 
$\lambda_\sigma = f^2$ and $\lambda_\sigma = g^2$, respectively (see Tab.~\ref{tab:classification}).

Apart from the operators summarised in Tab.~\ref{tab:classification}, 
we find the following four-fermion interactions:
\begin{align}
\mathcal{O}_{\psi\psi} &=(\overline{\psi} \gamma_\mu \psi) (\overline{\psi} \gamma^\mu \psi)\,, \label{eq:C2_D6_SMSM} \\ 
\mathcal{O}_{N\psi} &=(\overline{ N_R } \gamma_\mu  N_R ) (\overline{\psi} \gamma^\mu \psi)\,, \label{eq:C2_D6_nuSM} \\
\mathcal{O}_{\chi\psi} &=(\overline{\chi_L} \gamma_\mu \chi_L) (\overline{\psi} \gamma^\mu \psi)\,, \label{eq:C2_D6_DMSM}
\end{align}
where $\psi$ stands for the SM fermions, \textit{i.e.}~$\psi=L$, $e_R$, $Q$, $u_R$, $d_R$. The corresponding Wilson coefficients read:
\begin{equation}
\frac{c_{\psi\psi}}{\Lambda^2}=-\frac{g'^2 Q^2_\psi}{2m^2_{Z^\prime}}\,,\quad \frac{c_{N\psi}}{\Lambda^2}=-\frac{g'^2 Q_N Q_\psi}{m^2_{Z^\prime}}\,\quad \text{and} \quad \frac{c_{\chi\psi}}{\Lambda^2}=-\frac{g'^2 Q_\chi Q_\psi}{m^2_{Z^\prime}}\,.
\end{equation}
Here $Q_\psi$, $Q_N$ and $Q_\chi$ denote the $B-L$ charges 
of $\psi$, $N_R$ and $\chi_L$, respectively, see Tab.~\ref{tab:BLcharges}.
\begin{table}[t]
\centering
\begin{tabular}{l|ccccc|c|cc}
\toprule
 & $Q$ & $u_R$ & $d_R$ & $L$ & $e_R$ & $N^{1,2}_R$ & $\chi_L$ & $\sigma$   \\
\midrule
$U(1)_{B-L}$ & $+1/3$ & $+1/3$ & $+1/3$ & $-1$ & $-1$ & $-1$ & $+1$ & $+2$  \\
\bottomrule
\end{tabular}
\caption{$B-L$ charges of the particles in Model C2. $Q$ and $L$ are the SM quark and lepton $SU(2)_L$ doublets; $u_R$, $d_R$ and $e_R$ are the SM fermion singlets. 
$N_R^{1,2}$ are RH neutrinos, $\chi_L$ is a fermionic DM candidate, and 
$\sigma$ is a complex scalar.}
\label{tab:BLcharges}
\end{table}

The phenomenology of this model has been investigated in detail in Ref.~\cite{Escudero:2018fwn}, and its parameter space
has been shown to be severely constrained. 
For other studies exploring DM--neutrino connections 
in gauged $U(1)_{B-L}$ models, see
references in \cite{Escudero:2018fwn} 
as well as~\cite{FileviezPerez:2019cyn,Bandyopadhyay:2022xlp}.

\section{Phenomenology of selected renormalisable models} 
\label{sec:phenomenology}
%
In this section, we 
study the phenomenology of the UV completions presented in Sec.~\ref{sec:open}. The main 
focus of this work is 
on the neutrino portal regime, where the relic abundance is set by the DM annihilations into sterile neutrinos $\chi\chi \to NN$, with no connection with the SM through the Higgs or vector portals. However, we need some interaction  that guarantees the thermal equilibrium of $N$ with the SM particles in the early Universe, as has been mentioned in Sec.~\ref{sec:thermal_equi}. Therefore, we assume a (small) value for the Higgs portal coupling,  
$10^{-6} \lesssim \lambda_{\phi H} \lesssim 10^{-3}$, 
that keeps the dark sector in kinetic equilibrium with the SM 
up to a certain temperature (see App.~\ref{app:BEqs}).%
\footnote{A similar role can be played by the kinetic mixing for Models C.} 
For genuine Models~A, this coupling does not affect the DM relic abundance. 
However, in Models B2 and C2, when $\sigma$ and $H$ develop VEVs, 
the two scalars mix, and 
the coupling  $\lambda_{\sigma H}$ as small as $10^{-4}$--$10^{-3}$ 
would contribute significantly to the relic abundance around the resonance 
$m_\chi = m_h/2$, where $m_h$ is the mass of the physical Higgs boson, see \textit{e.g.} Ref.~\cite{Binder:2017rgn}. 
We will not discuss this effect in what follows. 

We summarise the main phenomenological features of the models described in the previous section in Tab.~\ref{tab:pheno}, while some further characteristics are described below.
\begin{itemize}
\item \textbf{Type-A models.} The Higgs portal term could produce DD signals at one loop. For $m_\chi<m_h/2$, this coupling would also induce invisible Higgs decay, $h\to\chi\chi$. 
In any case, if $\lambda_{\phi H} \ll 1$, these constraints are evaded.
Furthermore, there will always be a one-loop contribution to DD through the exchange of $Z$ boson. If $m_\chi<m_Z/2$, this will also lead to invisible $Z$ decay, $Z\to\chi\chi$. However, both processes are suppressed by the small neutrino Yukawa coupling. For a detailed analysis of such one-loop contributions see Ref.~\cite{Herrero-Garcia:2018koq}.
Model A2a has interesting features avoiding ID bounds because the DM annihilation cross section is $p$-wave, and in Model A2c finite $m_N$ 
is generated at one loop. We will discuss the latter in Sec.~\ref{sec:mnu_1loop}.
\item \textbf{Type-B models.} For Model B1, the most general scalar potential written in Eq.~\eqref{eq:B1_potential} includes the $\mu_{\phi H}$ term, that generates a VEV for the scalar $\phi$ upon EWSB. In that case, there is a mixing between the scalar and the Higgs, and  
elastic scattering of DM off nuclei occurs at tree level. In addition, if $m_\chi<m_h/2$ we also have invisible Higgs decay.
However, as we are interested in the  
neutrino portal regime, we take $\mu_{\phi H}=0$ in the phenomenological analysis; also the small value for the Higgs portal term does not generate any additional contribution to the relic abundance with processes involving SM particles, except for around the resonance $m_\chi = m_h/2$, 
if $\lambda_{\phi H} \gtrsim 10^{-4}$.

In Model B1, if the Yukawa coupling $g$ of DM to the scalar mediator 
is real, the annihilation cross section for $\chi\chi\to NN$ is $p$-wave. 
This is a reflection 
of the fact that DM requires a pseudo-scalar coupling (given by the imaginary part of $g$) to annihilate 
through $s$-wave, see \textit{e.g.} Ref~\cite{Berlin:2014tja}.
In Model B2, since both $f$ and $g$ are real, the annihilation cross section is also $p$-wave. Thus, in these cases, the
ID limits are avoided. 

Finally, this kind of models has four-fermion self-interactions of $N$ and $\chi$. However, the DM self-interactions $\chi\chi\leftrightarrow\chi\chi$ are 
very suppressed in the parameter space considered in our analysis, \textit{i.e.} $\sigma_{\chi\chi\to\chi\chi}/m_\chi \lesssim 10^{-6}\,\mathrm{cm^2/g}$ for Model B1, well below current limits~\cite{Tulin:2017ara}. 
\item \textbf{Type-C models.} Model C1 is not UV-complete and has to be understood as an effective description of the interaction between $N_R$ and $\chi_L$ via a new massive vector boson. Note that in the presence of kinetic mixing, there would be other processes like $Z\to\chi\chi$ if $m_\chi<m_Z/2$. Even if the tree-level parameter $\epsilon=0$, this kinetic mixing will be induced at one loop but will be further suppressed by the small neutrino Yukawa coupling.
Model C2, instead, leads to direct interactions of DM with the SM particles, cf. Eq.~\eqref{eq:C2_D6_DMSM}, and thus, it is severely 
constrained~\cite{Escudero:2018fwn}.
\end{itemize}
Given that the phenomenology of Models A1, B2 and C2 
have been studied in detail in Refs.~\cite{Escudero:2016ksa,Tang:2016sib,Bandyopadhyay:2018qcv,Batell:2017rol}, \cite{Escudero:2016tzx} and \cite{Escudero:2018fwn}, respectively, 
and that Model C1 is not UV-complete, we will analyse in detail Models A2 (b and c) and B1 in the next subsection. Moreover, in the last part we will comment on the one-loop generation of the RH neutrino mass in Model A2c.  
\begin{table}[t]
\centering
\begin{tabular}{l|c|ccc|cc|cc}
\toprule
\backslashbox{Feature}{Model} & A1 & A2a & A2b & A2c & B1 & B2 & C1 & C2 \\
\midrule
$s$-wave $\langle \sigma v \rangle_{\chi\chi \to NN}$ & \cmark & \xmark & \cmark & \cmark & * &\xmark & \cmark & \cmark \\
DD @ tree level & \xmark & \xmark & \xmark & \xmark & \cmark & \cmark & \xmark & \cmark \\
Self-interactions & \xmark & \xmark & \xmark & \xmark & \cmark & \cmark & \cmark & \cmark \\
\bottomrule
\end{tabular}
\caption{Classification of the phenomenological features of the UV completions discussed in Sec.~\ref{sec:open}. Notice that when the DM annihilation cross section $\langle \sigma v \rangle_{\chi\chi\to NN}$ is $p$-wave, one can easily avoid indirect bounds. The asterisk * in Model B1 implies that if the coupling $g$ is real, the thermally averaged DM annihilation cross section is $p$-wave. 
}\label{tab:pheno}
\end{table}
%

\subsection{Dark matter phenomenology} 
\label{sec:DM}
%
In this subsection, we focus on models A2b, A2c and B1. 
As discussed in Sec.~\ref{sec:A2}, for one generation of $N_R$ and $\chi_{L}$, the coupling $f$ in Models A2b and A2c can be rendered real.
On the other hand, for Model B1 both couplings $f$ and $g$ are complex, in general, \textit{i.e.}~$f=f_r+if_i$ and $g=g_r+ig_i$. Therefore,
we consider two 
cases: (i) real $f$ and $g$, and (ii) purely imaginary $f$ and $g$. 
In the former, CP-conserving case, 
DM annihilates into sterile neutrinos via $p$-wave, evading ID limits.

The general expressions for the coefficients $a$ and $b$ in Eq.~\eqref{eq:Xsec_thermal} for the thermally averaged DM annihilation cross section $\chi\chi\to NN$ are given in what follows, where we use that $r_i \equiv m_i/m_\chi$. 
\begin{itemize}
\item \textbf{Model A2b}, with coupling $f$ real:
\begin{equation}
a =\frac{f^4}{16 \pi  m_{\chi }^2} 
\frac{r_N^2 \sqrt{1-r_N^2}}{\left(1+r_{\sigma }^2-r_N^2\right)^2}\,.
\label{eq:A2b_realf}
\end{equation}
\item \textbf{Model A2c}, with coupling $f$ real:
\begin{equation}
a =\frac{f^4}{64\pi m_{\chi}^{2}} \frac{\sqrt{1-r_N^{2}} 
\left(r_\rho^2 - r_{\theta}^{2} 
- \left(2 + r_\rho^2 + r_{\theta}^{2}\right) r_N + 2r_N^{3}\right)^{2}}
{\left(1+r_{\rho}^{2}-r_N^{2}\right)^{2}\left(1+r_{\theta}^{2}-r_N^{2}\right)^{2}}
\,.
\label{eq:A2c_realf}
\end{equation}
\item \textbf{Model B1}, with couplings $f$ and $g$ complex:
\begin{align}
 a &= \frac{4 g_i^2}{\pi m_\chi^2} \frac{\sqrt{1-r_N^2}}{\left(r_\phi^2 - 4\right)^2}
 \left[f_i^2 + f_r^2 \left(1 - r_N^2\right)\right], \\
 b &= \frac{2}{\pi m_\chi^2 \left(r_\phi^2-4\right)^3 \sqrt{1 - r_N^2}}\, 
   \bigg\{f_i^2 \left[g_i^2 \left(16 + r_N^2 \left(r^2_\phi - 20\right)\right) 
   + 2 g_r^2 \left(1 - r_N^2\right) \left(r_\phi^2-4\right)\right] \nonumber \\
   &\hspace{2.8cm} +f_r^2 \left(1 - r_N^2\right)
   \left[g_i^2 \left(16 + r_N^2 \left(3r^2_\phi - 28\right)\right)
   + 2 g_r^2 \left(1 - r_N^2\right) \left(r_\phi^2-4\right)\right]\bigg\}\,.
\end{align}
We consider the following two situations: 
\subitem (i) 
Real $f$ and $g$, for which 
\begin{equation}
a = 0\,,\quad
b = \frac{4 f^2_r g^2_r}{\pi m_{\chi}^2 } 
\frac{\left(1-r_N^2\right)^{3/2}}{\left(r_{\phi }^2-4\right)^2}\,.
\label{eq:B1realfg}
\end{equation}
\subitem (ii)  
Purely imaginary $f$ and $g$, resulting in
\begin{equation}
a =\frac{4 f^2_i g^2_i}{\pi m_{\chi}^2 } 
\frac{\sqrt{1-r_N^{2}}}{\left(r_{\phi }^2-4\right)^2}\,.
\label{eq:B1imaginaryfg}
\end{equation}
\end{itemize}
We see that, in all the cases except Model B1 with real couplings, 
the DM annihilation cross section is $s$-wave. 
\begin{figure}[t!]
 \centering
 \includegraphics[width=.49\textwidth]{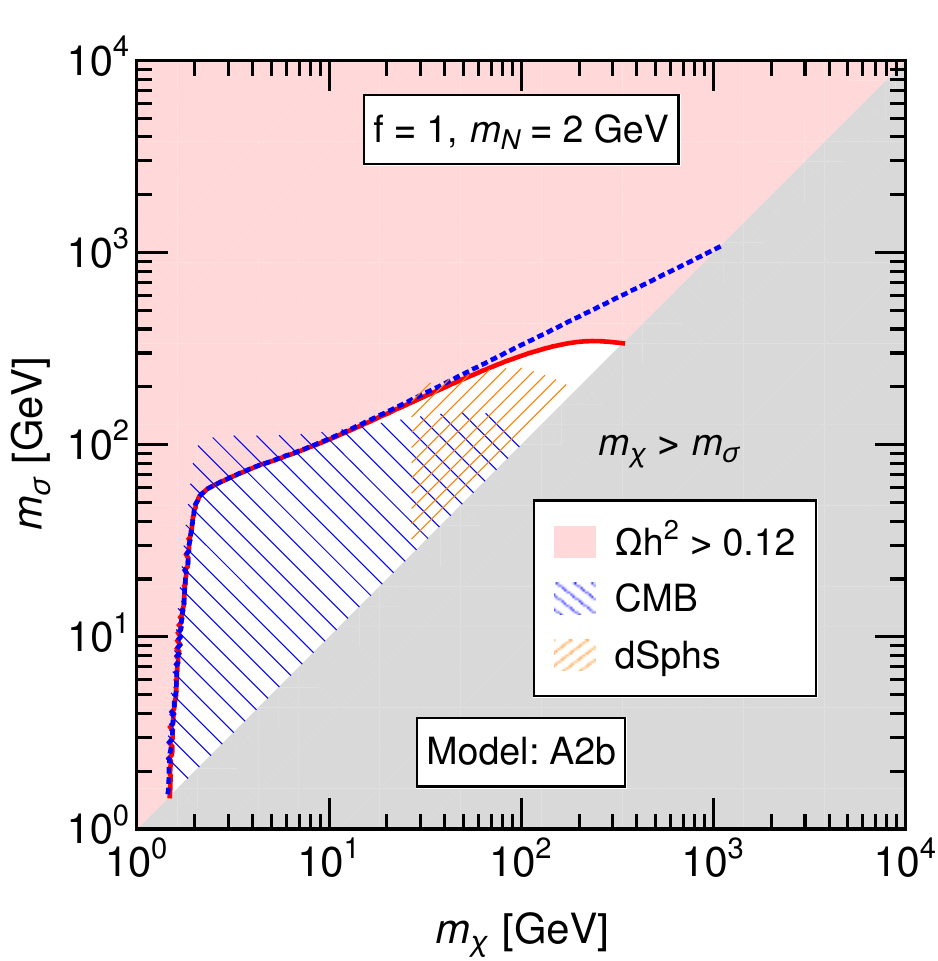}
 \includegraphics[width=.49\textwidth]{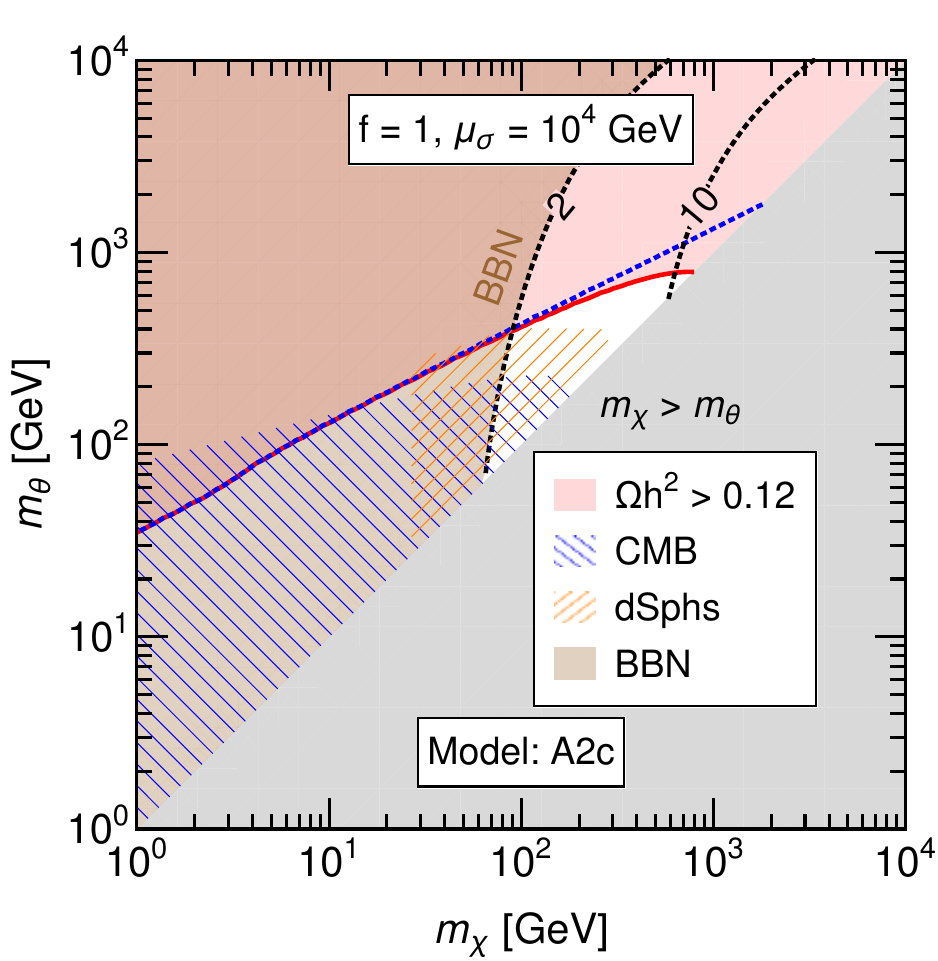}
 \includegraphics[width=.49\textwidth]{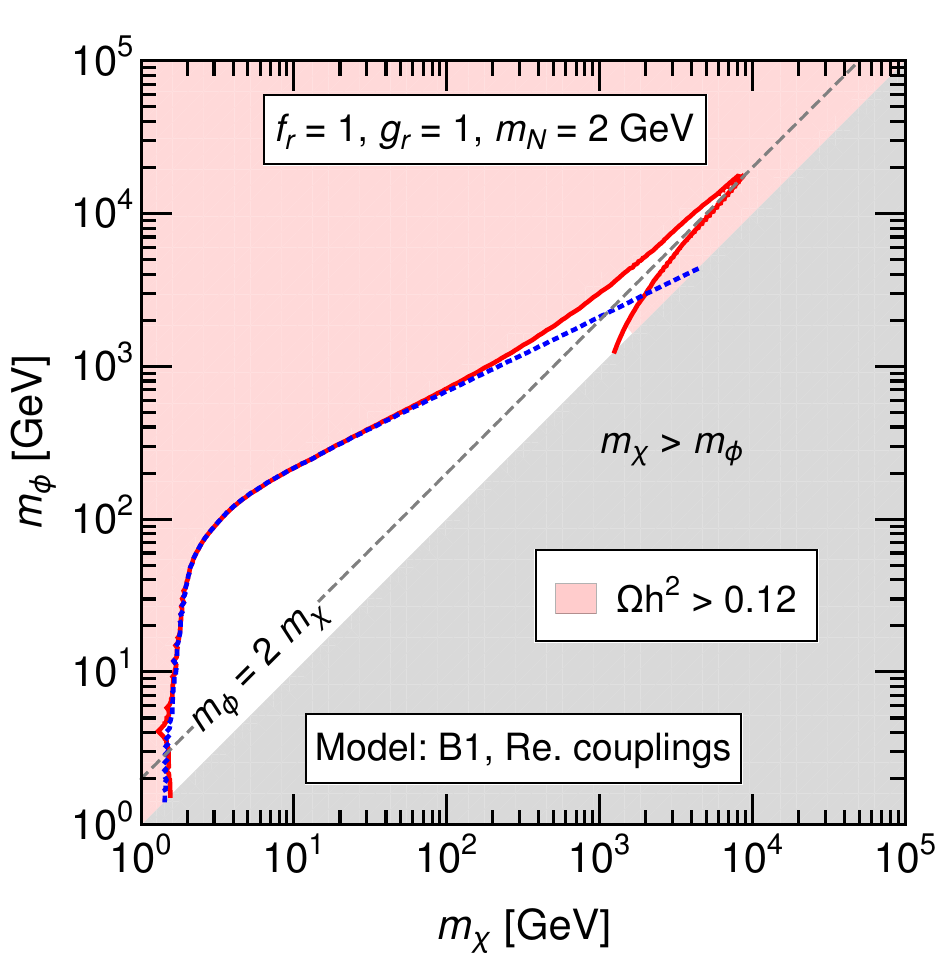}
 \includegraphics[width=.49\textwidth]{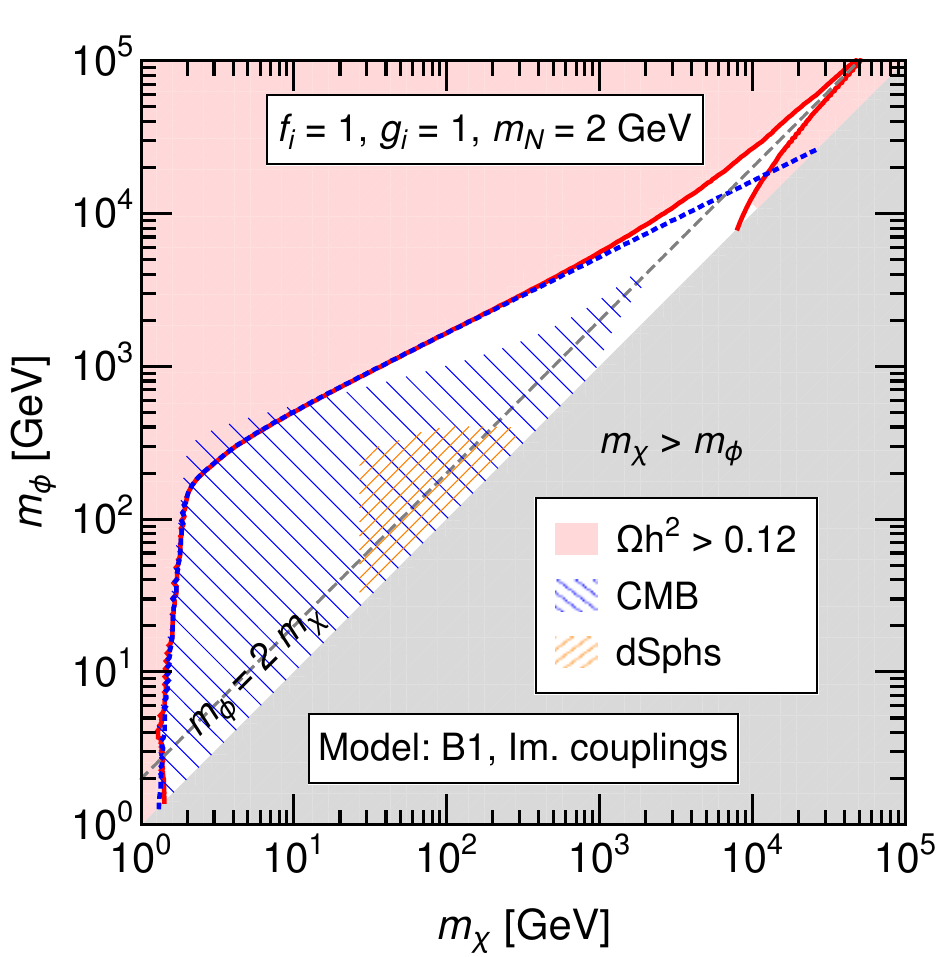}
 \caption{Relevant parameter space 
 of models A2b (top left), A2c (top right) and 
 B1 with real (bottom left) and imaginary (bottom right) couplings. 
 For each model, the values of the fixed parameters are specified in the upper region of the plots. 
 Along the red line, the observed value of DM relic abundance is reproduced. 
 The blue dotted line corresponds to the calculation of relic abundance in the EFT approach.
 When applicable, we present experimental bounds from ID 
 and BBN. See details in the text.}
 \label{fig:Results_plot}
\end{figure}

In Fig.~\ref{fig:Results_plot}, we depict the thermally averaged 
DM annihilation cross section corresponding to the observed value 
of DM relic abundance $\Omega h^2 = 0.12$~\cite{Planck:2018vyg} 
using the results from Ref.~\cite{Bringmann:2020mgx}.%
\footnote{See Figs.~1 and~4 in this reference for the $s$- and $p$-wave DM annihilation cross section, respectively.}
The cross section is computed both (i) in a renormalisable model 
(red line) and (ii) in the corresponding EFT (blue dotted line), 
using the matching conditions from Tab.~\ref{tab:classification}. 
Therefore, red and blue dotted lines correspond to the values of  
$\langle \sigma v \rangle$ that reproduce the observed relic abundance, as explained before.
The four panels correspond to Models A2b (top left), A2c (top right) and B1 with real (bottom left) and purely imaginary (bottom right) Yukawa couplings.\footnote{As pointed in Ref.~\cite{Gondolo:1990dk}, the expansion in Eq.~\eqref{eq:Xsec_thermal} fails in some scenarios, in particular, near thresholds and resonances. In those cases, for $\langle \sigma v \rangle$ we use the full expression in Eq.~\eqref{eq:thermalsigmav_Gondolo}.} We observe that the EFT approach to the calculation of DM relic abundance works 
for $m_\chi \lesssim m_{\sigma,\theta}/4$ in the type-A models 
and $m_\chi \lesssim m_\phi/6$ in Model B1. For the latter, the resonance behaviour of the cross section clearly cannot be captured by the EFT.
Our analytical results for the relic abundance 
agree with the results obtained using \texttt{micrOMEGAs}~\cite{Belanger:2001fz,Belanger:2018ccd}. Regions in red stand for values of the relic abundance which would overclose the Universe, \textit{i.e.} $\Omega h^2>0.12$. 
We show the values of the parameters that are fixed for each model  
in the upper region of the plots, taking for the RH neutrino mass in models A2b and B1 the minimal value allowed by the BBN constraint, \textit{i.e.} $m_N=2\,\gev$. As we will discuss in Sec.~\ref{sec:mnu_1loop}, in Model A2c the sterile neutrino mass $m_N$ is generated radiatively. In Fig.~\ref{fig:Results_plot}, we take $\mu_\sigma=10^4\,\gev$ in order 
to have $m_N \gtrsim 2$~GeV in 
the part of the parameter space of interest, and contours of fixed $m_N$ (in GeV) are also shown as black dotted lines. 
In addition, the brown region is excluded by the BBN constraint. We assume $f=1$ as an illustrative example for models A2b and A2c without loss of generality. For Model B1, we distinguish between the cases of real and purely imaginary couplings, $f_r=g_r=1$ and $f_i=g_i=1$, respectively.

We also add ID constraints from Ref.~\cite{Batell:2017rol}, which we  briefly discuss below. \textit{Planck} cosmic microwave background (CMB) measurements set bounds on the DM annihilations into SM particles. The related production of particles leads to homogenisation of the CMB power spectra and the modification of the ionisation history of the Universe. In addition, the \textit{Fermi} analysis of dwarf spheroidal galaxies (dSphs) provides limits on the DM annihilation cross section by non-observation of excess above the astrophysical backgrounds in the gamma-ray flux, with photon energies in the $500\,\mev$--$500\,\gev$ range.
In the models discussed here, DM has negligible interactions with the SM particles, in particular with quarks. Therefore, it is not captured in the Sun and no associated ID constraints exist.
The bounds from CMB and dSphs are represented in the plots by the blue and orange hatched regions, respectively. Notice that for Model B1 
in the case of real $g$ (coupling of DM to the scalar mediator), indirect bounds do not apply due to the $p$-wave nature of the annihilation cross section $\chi\chi\to NN$.

Finally, for the parameters of the models that we take in Fig.~\ref{fig:Results_plot}, the white regions correspond to points that avoid all the experimental bounds and provide some fraction of the total relic abundance of DM. For Model A2b, 
a small region of the parameter space with 
$100\,\gev \lesssim m_\chi \lesssim 300\,\gev$ 
and $200\,\gev \lesssim m_\sigma \lesssim 300\,\gev$ is open. 
In Model A2c, DM masses between approximately $100\,\gev$ and $800\,\gev$, 
in conjunction with $300\,\gev \lesssim m_\theta \lesssim 800\,\gev$ are allowed. In this case, the values for the one-loop generated RH neutrino mass are $2\,\gev \lesssim m_N \lesssim 10\,\gev$. For Model B1, due to the resonance behaviour of the annihilation cross section, 
a larger part of the parameter space 
is open. For real couplings, $\chi$ with mass between $2$~GeV and $10$~TeV, and $\phi$ with mass between $2$~GeV and $20$~TeV can be responsible for the totality of observed DM relic abundance, 
whereas for purely imaginary couplings, the allowed interval of DM masses 
is $30\,\gev \lesssim m_\chi \lesssim 50\,\tev$ 
and that of the scalar mediator masses is $1\,\tev \lesssim m_\phi \lesssim 100\,\tev$.

Differences in the results shown in Fig.~\ref{fig:Results_plot} could, in principle, come from allowing other particles of the dark sector to evolve out of the thermal equilibrium and enabling particles to decay, as it is detailed in App.~\ref{app:BEqs} for Model A2b. However, the evolution of the full set of Boltzmann equations shows that the deviations are not significant in almost all of the parameter space analysed in this work.

\subsection{Neutrino masses in Model A2c} 
\label{sec:mnu_1loop}
%
The Lagrangian of the model is given in Eq.~\eqref{eq:LagA2c}.
The $U(1)_L$ symmetry, under which the complex scalar $\sigma$ 
has charge (+1), is softly broken by a quadratic term in the potential, 
whereas $m_N = 0$ at tree level. The soft breaking term splits the masses 
of the real, $\rho$, and imaginary, $\theta$, components of $\sigma$, 
see Eq.~\eqref{eq:masses}.
We choose $\mu_\sigma^2 > 0$, such that $m_\rho > m_\theta$. The lighter of the $Z_2$-odd fields, \textit{i.e.} either $\chi$ or $\theta$, yields a DM candidate. However, here we only focus on fermionic DM. 
We notice that other $U(1)_L$ breaking terms are possible, as \textit{e.g.} 
$\lambda'_{\sigma H} \sigma^2 |H|^2$  or $m_N \overline{N_R^c}  N_R$, 
but they are \emph{harder} (higher dimension). Even if these terms are absent at tree level, finite contributions to both of them are generated at one loop. 
In the case of $ \lambda'_{\sigma H}$, this reads
\begin{equation}
 \lambda'_{\sigma H} \simeq \frac{ \lambda_{\sigma H} \,\lambda_\sigma\,  \mu_{\sigma}^2}{16\pi^2 m_\sigma^2}\,. \label{lambdasoft}
\end{equation}

\begin{figure}[t]
 \centering
 \includegraphics[width=.8\textwidth]{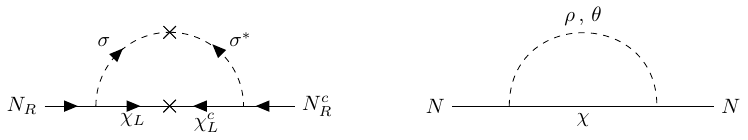}
 \caption{Radiative generation of $m_N$ in Model A2c. 
On the left, the diagram is drawn in the interaction basis, 
whereas on the right, in the mass basis. 
The crosses denote mass insertions $\mu_\sigma^2$ and $m_\chi$.}
 \label{fig:ISS}
\end{figure}

The splitting in the masses of $\rho$ and $\theta$ 
leads to a finite $m_N$ being generated at one loop, see Fig.~\ref{fig:ISS}. A similar mechanism has been proposed in Ref.~\cite{Ahriche:2016acx}. 
It is analogous to that of the scotogenic model~\cite{Ma:2006km} and its generalisations~\cite{Hagedorn:2018spx,Beniwal:2020hjc,Escribano:2020iqq}, where light (mostly-active) neutrinos acquire their mass through one-loop diagrams, for a review see Ref.~\cite{Cai:2017jrq}. After EWSB the tree-level Majorana mass term for electrically neutral fermions reads 
\begin{equation}
  \mathcal{L}_{\mathrm{A2 c}} \supset -\frac{1}{2}
 \begin{pmatrix} \overline{\nu_L} & \overline{ N^c_R } & \overline{\chi_L} \end{pmatrix} 
 \begin{pmatrix} 
0 & m_{\rm D} & 0 \\
m^T_{\rm D} & 0 & 0 \\
0 & 0 & m_\chi \\
\end{pmatrix}
\begin{pmatrix} \nu^c_L \\ N_R \\ \chi^c_L \end{pmatrix} + \text{H.c.} \,,\label{eq:mnu_tree}
\end{equation}
where $m_{\rm D}=y_\nu v_h/\sqrt{2}$. 
The DM candidate $\chi$ is decoupled from the rest of neutral leptons, 
since it is charged under the $Z_2$.

For the computation of $m_N$ at one loop, we assume that there are 
$n_N$ generations of $N_R$ and $n_\chi$ generations of $\chi_L$. 
Furthermore, we work in a basis for $\chi_L$ in which $m_\chi$ is diagonal with positive and real elements $m_{\chi_k}$.
Performing the computation (which is identical to the one in the scotogenic model), we find
\begin{equation}
 \left(m_N\right)_{ij} =\sum_{k = 1}^{n_\chi} 
 \frac{ f_{ik}^\ast f_{jk}^\ast m_{\chi_k}}{32 \pi^2} 
 F\left(m_\rho^2, m_\theta^2, m_{\chi_k}^2\right),
 \label{eq:mRloop}
\end{equation}
where the loop function $F$ is defined as follows:%
\begin{equation}
 F(x, y, z) = \frac{x}{x-z}\log\left(\frac{x}{z}\right) - \frac{y}{y-z}\log\left(\frac{y}{z}\right). 
 \label{eq:Floop}
\end{equation}
From this formula, it is evident that in the limit  
of lepton number conservation 
($\mu_\sigma^2 =0$, and hence, $m_\rho = m_\theta$), 
the mass matrix $m_N = 0$. 
In fact, in the limit $m_{\chi_k}\ll m_\rho, m_\theta$ and 
$m_\rho \simeq m_\theta \simeq m_\sigma$ one finds
\begin{equation}
 \left(m_N\right)_{ij} \approx \frac{\mu_\sigma^2}{16 \pi^2 m_\sigma^2} \sum_{k = 1}^{n_\chi} 
 f_{ik}^\ast f_{jk}^\ast m_{\chi_k} .
 \label{eq:mRloop-approx}
\end{equation}
This result, which depends linearly on $m_{\chi_k}$, can easily be estimated by using dimensional analysis and symmetry arguments.

Depending on the number of generations $n_N$ and $n_\chi$, 
some of the RH neutrinos may remain massless. For instance, 
if $n_N$ = 3 and $n_\chi = 1$, only one of the eigenvalues of $m_N$ is non-zero, since $m_N \sim f^\ast f^\dagger$ in this case. 
$n_\chi = 2$ leads to two massive $N$, 
a minimal number (within the type I seesaw mechanism) 
needed to explain low-energy neutrino oscillation data. 
For $n_\chi = 3$, all three RH neutrinos get masses. In the case of only one generation $n_N=n_\chi=1$, and assuming $m_N \gg m_{\rm D}$, we can express the coupling $f$ as 
\begin{equation}
f=4\pi \frac{y_\nu\,v_h}{\sqrt{m_\nu\,m_\chi\,F(m^2_\rho,m^2_\theta,m^2_\chi)}}\,,\label{eq:f_coupling}
\end{equation}
with $m_\nu \sim 0.05$~eV the mass of the light active neutrino and $v_h=246\,\gev$ the Higgs VEV.

\section{Summary and conclusions}
\label{sec:conc}
%
Motivated by the lack of WIMPs signals, in the present work we have revisited the possibility 
of SM singlet DM interacting primarily with sterile neutrinos.
The latter can explain the light neutrino masses.
We have extended the SM with 
a Majorana fermion $\chi$ and RH neutrinos $N_R$, 
assuming that their interactions are described by effective four-fermion operators. 
The stability of $\chi$ is ensured by a $Z_2$ symmetry. 
Restricting ourselves to the case in which DM interacts with the lightest of sterile neutrinos, we have shown that there are three independent four-fermion operators.
One of them, $\mathcal{O}_1$, always preserves lepton number, 
whereas the remaining ones, $\mathcal{O}_2$ and $\mathcal{O}_3$, 
may either preserve or violate it 
(depending on the lepton number of $\chi_L$).
We refer to $\mathcal{O}_{1,2,3}$ as \textit{sterile neutrino portal operators}.

Assuming that the mass of the lightest sterile neutrino, $m_N$, 
is smaller than that of $\chi$, the observed DM relic abundance 
can be entirely explained by the freeze-out of $\chi$ 
due to the annihilation process $\chi \chi \to N N$ triggered by the portal operators. 
For $\mathcal{O}_1$, the $s$-wave part 
of the corresponding annihilation cross section is proportional to $m_N^2$, 
and thus, is suppressed for small values of $m_N$. 
For $\mathcal{O}_2$ and $\mathcal{O}_3$, there is no such a suppression. 
Turning one operator at a time we have derived the scale of new physics $\Lambda$ 
required to reproduce the observed relic abundance.

Further, we have formulated simple UV completions that lead 
to one or more portal operators when integrating out a heavy mediator at tree level. 
Depending on the Lorentz nature of the mediator and whether it propagates 
in $t$- or $s$-channel of the $\chi \chi \to N N$ process, 
we classified the UV models into
\begin{itemize}
 \item Model A1 (A2) containing a real (complex) scalar $\phi$ ($\sigma$) propagating in $t$-channel;
 \item Model B1 (B2) involving a real (complex) scalar $\phi$ ($\sigma$) propagating in $s$-channel;
 \item Model C1 (C2) having a massive vector (gauge) boson $Z'$
  propagating in $s$-channel.%
 \footnote{Since Model C2 is a gauged version of Model B2, it also involves a complex scalar.}
\end{itemize}
In Models A, the mediator is charged under the $Z_2$ symmetry 
stabilising DM, whereas in Models B and C, the mediators are neutral 
under this symmetry.
For Model A2, we have considered three different situations: 
(i) $m_N = 0$ and $U(1)_L$ is conserved (A2a), corresponding to light neutrinos being Dirac particles; 
(ii) $m_N \neq 0$ with the scalar potential preserving lepton number (A2b); 
and (iii) $m_N = 0$, but $U(1)_L$ being softly broken by the $\mu_\sigma^2 \sigma^2$ in the potential.
Model B2 (C2) possesses the global (local) $U(1)_{B-L}$ symmetry,  
spontaneously broken in two units by the VEV of the complex scalar.
Instead, Model C1 should be viewed as a low-energy effective description of the interaction
mediated by a massive vector boson. 

For each of the models in the list above, we have worked out 
the corresponding EFT operators invariant under the SM gauge symmetry 
(and $Z_2$ stabilising DM) to dimension six. 
We have found that in Models A1 and A2, 
the only effective interactions generated at $D \leq 6$ 
are the neutrino portal operators $\mathcal{O}_1$ and $\mathcal{O}_2$. 
We have dubbed these model \textit{genuine}.
On the contrary, in addition to the portal operators $\mathcal{O}_2$ and $\mathcal{O}_3$, 
Models B1 and B2 induce $D=5$ interactions of $\chi$ and $N$ with the Higgs boson. 
The Wilson coefficients of these operators are proportional to the  
coupling $\mu_{\phi H}$ ($\lambda_{\sigma H}$) in the scalar potential for Model B1 (B2). 
Thus, if these couplings are sufficiently small, 
the DM phenomenology of 
Models B can be dominated by the neutrino portal operators. Moreover, we find four-fermion self-interactions 
of $\chi$ and $N$ controlled by the same Yukawa couplings 
that enter the matching relations for $c_2$ and $c_3$. 
In addition, the EFT of Model B2 contains a (massless) Goldstone boson, 
the Majoron, and its derivative $D=5$ interactions with all fermions 
carrying non-zero lepton number as well as a $D=6$ interaction with the Higgs boson.  
Finally, while the effective Model C1 leads only to $\mathcal{O}_1$ and the self-interactions of $\chi$ and $N$, the gauged $U(1)_{B-L}$ model (Model C2) gives rise 
to all three portal operators,  
$D=5$ interactions of $\chi$ and $N$ with the Higgs%
\footnote{As in Model B2, these interactions are controlled by an independent coupling 
$\lambda_{\sigma H}$ in the potential.}
and four-fermion operators involving $\chi$, $N$ and/or the SM fields.

In Model A2a, where $m_N = 0$ and light neutrinos are Dirac, 
the annihilation cross section is  
effectively $p$-wave, and 
the ID bounds from annihilation to neutrinos~\cite{Arguelles:2019ouk} are avoided. 
This scenario is interesting since it allows for light thermal DM, 
with masses as small as $100$~MeV. 
However, it is very difficult to probe it.  
In Models A2b and A2c, $m_N \neq 0$ at tree and one-loop level, respectively. Model A2c possesses this interesting feature of finite $m_N$ being generated at 
one-loop level, analogously to the generation of light neutrino masses 
in the scotogenic model. 
In the limit of DM mass being smaller than the mass of the scalar mediators, $m_N \sim m_\chi/(4\pi)^2$.
In other models considered in the present work, 
$m_N$ is a free parameter. 
In any case, it should be larger than approximately 2 GeV for $N$ to decay before BBN. 
Decays of $N$ will modify the spectra of charged particles (in particular, antiprotons and positrons) as well as photons. These modifications can be looked for in ID
as discussed in Ref.~\cite{Batell:2017rol}. We have adopted the constraints from 
CMB measurements by Planck and dSphs observations by Fermi derived in Ref.~\cite{Batell:2017rol}, 
showing that in Model A2b (A2c) a Majorana fermion $\chi$ with $m_\chi$ between approximately 
100~GeV and 300~GeV (800~GeV) can constitute 100\,\% of the observed DM relic abundance, respecting the ID constraints. 

In Model B1, the annihilation cross section is $p$-wave 
if the Yukawa coupling of DM to a scalar mediator, $g$, is real.
In this case, the ID bounds are avoided, 
and DM masses between 2~GeV and 10~TeV are allowed. 
On the contrary, for complex $g$, the annihilation cross section is $s$-wave, 
and thus, the ID constraints apply. 
For purely imaginary $f = g = i$, we find that the viable range of 
DM masses is $30\,\gev \lesssim m_\chi \lesssim 50\,\tev$.  
Larger DM masses in this model are accessible due to the resonance enhancement 
of the cross section. 

In conclusion, we have shown that 
DM--sterile neutrino interactions 
described by effective four-fermion operators
constitute a viable option. 
They can be generated in a number of UV-complete models 
possessing somewhat distinct phenomenology. 
This scenario provides
a possible connection between neutrinos and dark matter, 
which arguably are among the most feebly interacting sectors of nature.

\section*{Acknowledgements}
%
This work is partially supported by the FEDER/MCIyU-AEI grant FPA2017-84543-P, 
the MICINN/ AEI (10.13039/501100011033) grants PID2020-113334GB-I00 and PID2020-113644GB-I00, 
  and the ``Generalitat Valenciana" grant PROMETEO/2019/087.
 JHG and CF are supported by the ``Generalitat Valenciana" through the GenT Excellence Program (CIDEGENT/2020/020). 
 LC and CF are also supported by the ``Generalitat Valenciana" under the ``GRISOLIA" and ``ACIF" fellowship programs, respectively.

\appendix
\section{Boltzmann equations: comparison to the standard approach} 
\label{app:BEqs}
%
%
\subsection{Departure from chemical equilibrium within the dark sector}
%
In this appendix we review the comparison of calculating the relic abundance with the full set of Boltzmann equations (BEqs) and the standard approximation (STD). We denote as STD the case considering just the evolution of the DM particle $\chi$ and the $2 \leftrightarrow 2$ processes. For example, in Model A2b described in Sec.~\ref{sec:A2}, the complete BEqs for $m_\theta = m_\rho > m_\chi,m_N$, can be expressed as:
\begin{align}
x \, s \, H \frac{dY_N}{dx} & = 
 \langle\sigma v\rangle_{\chi \chi \rightarrow N N}\,s^2 \left( Y_\chi^2 - \left(\frac{Y_\chi^{\mathrm{eq}}}{Y_N^{\mathrm{eq}}}\right)^2 Y_N^2 \right) + \langle\sigma v\rangle_{\theta \theta \rightarrow N N}\,s^2 \left( Y_{\theta}^2 -\left(\frac{Y_\theta^{\mathrm{eq}}}{Y_N^{\mathrm{eq}}}\right)^2 Y_N^2 \right) \nonumber \\
  & + \langle\sigma v\rangle_{\rho \rho \rightarrow N N}\,s^2 \left( Y_{\rho}^2 -\left(\frac{Y_\rho^{\mathrm{eq}}}{Y_N^{\mathrm{eq}}}\right)^2 Y_N^2 \right) - s \, \tilde{\Gamma}_{N} \left(Y_N - Y_N^{\mathrm{eq}}\right) \nonumber \\
  & + s \, \tilde{\Gamma}_{\theta} \left( Y_{\theta} - \frac{Y_{\chi}Y_{N}Y_\theta^{\mathrm{eq}}}{Y_\chi^{\mathrm{eq}}Y_N^{\mathrm{eq}}} \right) + s \, \tilde{\Gamma}_{\rho} \left( Y_{\rho} - \frac{Y_{\chi}Y_{N}Y_\rho^{\mathrm{eq}}}{Y_\chi^{\mathrm{eq}}Y_N^{\mathrm{eq}}} \right) \nonumber \\
  & + \langle\sigma v\rangle_{\theta \rho \rightarrow N N}\,s^2 \left( Y_\theta Y_\rho - Y_\theta^{\mathrm{eq}} Y_\rho^{\mathrm{eq}} \left(\frac{Y_{N}}{Y_N^{\mathrm{eq}}} \right)^2 \right) \,, \label{eq:BEqs_1} \\
x \, s \, H \frac{dY_\chi}{dx} & = -\langle\sigma v\rangle_{\chi \chi \rightarrow N N}\,s^2 \left( Y_\chi^2 - \left(\frac{Y_\chi^{\mathrm{eq}}}{Y_N^{\mathrm{eq}}}\right)^2 Y_N^2 \right) + 
  \langle\sigma v\rangle_{\theta \theta \rightarrow \chi \chi}\,s^2 \left( Y_{\theta}^2 -\left(\frac{Y_\theta^{\mathrm{eq}}}{Y_\chi^{\mathrm{eq}}}\right)^2 Y_\chi^2 \right) \nonumber \\
  & +  \langle\sigma v\rangle_{\rho \rho \rightarrow \chi \chi}\,s^2 \left( Y_{\rho}^2 -\left(\frac{Y_\rho^{\mathrm{eq}}}{Y_\chi^{\mathrm{eq}}}\right)^2 Y_\chi^2 \right) + s \, \tilde{\Gamma}_{\theta} \left( Y_{\theta} - \frac{Y_{\chi}Y_{N}Y_\theta^{\mathrm{eq}}}{Y_\chi^{\mathrm{eq}}Y_N^{\mathrm{eq}}} \right) + s \, \tilde{\Gamma}_{\rho} \left( Y_{\rho} - \frac{Y_{\chi}Y_{N}Y_\rho^{\mathrm{eq}}}{Y_\chi^{\mathrm{eq}}Y_N^{\mathrm{eq}}} \right) \nonumber \\
  & + \langle\sigma v\rangle_{\theta \rho \rightarrow \chi \chi}\,s^2 \left( Y_\theta Y_\rho - Y_\theta^{\mathrm{eq}} Y_\rho^{\mathrm{eq}} \left(\frac{Y_\chi}{Y_\chi^{\mathrm{eq}}} \right)^2 \right) \,, \label{eq:BEqs_2} \\
x \, s \, H \frac{dY_\theta}{dx} & = -\langle\sigma v\rangle_{\theta \theta \rightarrow N N}\,s^2 \left( Y_{\theta}^2 -\left(\frac{Y_\theta^{\mathrm{eq}}}{Y_N^{\mathrm{eq}}}\right)^2 Y_N^2 \right) - \langle\sigma v\rangle_{\theta \theta \rightarrow \chi \chi}\,s^2 \left( Y_{\theta}^2 -\left(\frac{Y_\theta^{\mathrm{eq}}}{Y_\chi^{\mathrm{eq}}}\right)^2 Y_\chi^2 \right) \nonumber \\
  & - s \, \tilde{\Gamma}_{\theta} \left( Y_{\theta} - \frac{Y_{\chi}Y_{N}Y_\theta^{\mathrm{eq}}}{Y_\chi^{\mathrm{eq}}Y_N^{\mathrm{eq}}} \right) - \frac{1}{2}\langle\sigma v\rangle_{\theta \rho \rightarrow \chi \chi}\,s^2 \left( Y_\theta Y_\rho - Y_\theta^{\mathrm{eq}} Y_\rho^{\mathrm{eq}} \left(\frac{Y_\chi}{Y_\chi^{\mathrm{eq}}} \right)^2 \right) \nonumber \\
  & -\frac{1}{2}\langle\sigma v\rangle_{\theta \rho \rightarrow N N}\,s^2 \left( Y_\theta Y_\rho - Y_\theta^{\mathrm{eq}} Y_\rho^{\mathrm{eq}} \left(\frac{Y_{N}}{Y_N^{\mathrm{eq}}} \right)^2 \right)\,, \label{eq:BEqs_3} \\
  x \, s \, H \frac{dY_\rho}{dx} & = x \, s \, H \frac{dY_\theta}{dx}\, [\theta \longleftrightarrow \rho]\,,\label{eq:BEqs_4}
\end{align}
in terms of the yields $Y_i=n_i/s$, where $n_i$ is the number density for species $i$ and $s$ is the total entropy density, $x=m_\chi/T$, $H$ is the ($x$-dependent) Hubble rate, and the superscript ``eq'' denotes equilibrium distributions with zero chemical potential, as in Refs.~\cite{Gondolo:1990dk,Kolb:1990vq}. 
$\tilde{\Gamma}_i$ are the thermal decay rates for a decaying particle $i$ given by
\begin{equation}
\tilde{\Gamma}_{i} = \frac{1}{n^{\mathrm{eq}}_i} \int \frac{d^3p_i}{\left(2\pi \right)^3 E_i} f^{\mathrm{eq}}_i m_i \Gamma_i\,,
\end{equation}
and $\Gamma_i$ are the zero-temperature decay rates. When $m_N<m_h\left(T \right)$, decays into $N$ should be taken into account by means of the following substitution $\tilde{\Gamma}_{N} \rightarrow \tilde{\Gamma}_{h}$ with,
\begin{equation}
\tilde{\Gamma}_{h} = \frac{1}{n^{\mathrm{eq}}_N} \int \frac{d^3p}{\left(2\pi \right)^3 E_h} f^{\mathrm{eq}}_h m_h \Gamma_h\,.
\end{equation}
Here we consider the decay with the approximation as in Ref.~\cite{Hambye:2016sby}, where all the four states of the Higgs doublet have the Higgs boson mass $m_h\left(T \right)$; thermal masses were taken from~\cite{Tang:2016sib}.
The thermally averaged cross section $\langle\sigma v\rangle$ is given by~\cite{Gondolo:1990dk}:
\begin{equation}
\langle\sigma v\rangle = \frac{1}{8m^4TK_2^2\left(m/T\right)} \int_{4m^2}^{\infty} ds\, \sigma (s) \left[ s-4m^2\right] \sqrt{s} \, K_1\left(\sqrt{s}/T\right), \label{eq:thermalsigmav_Gondolo}
\end{equation}
where $K_{1,2}$ are modified Bessel functions, and $m$ is the DM mass.

\begin{figure}[t]
 \centering
 
\includegraphics[width=.49\textwidth]{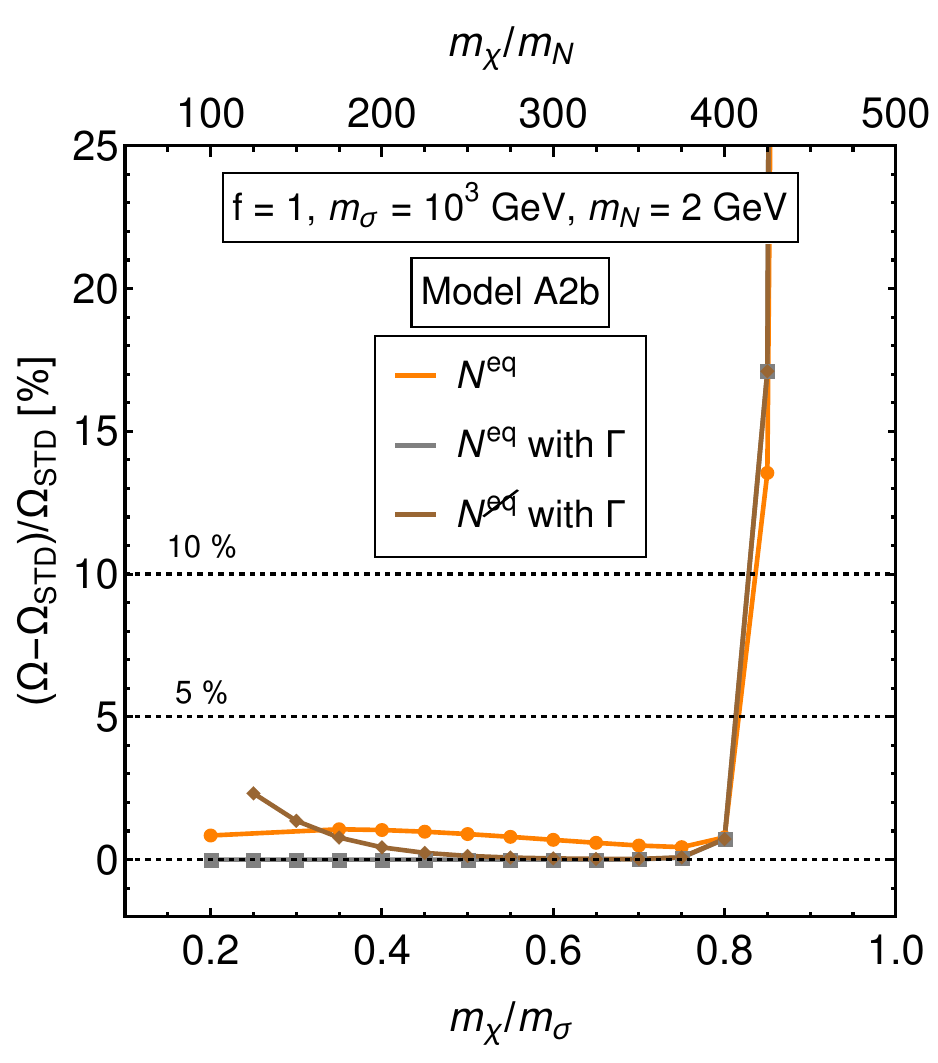}\includegraphics[width=.49\textwidth]{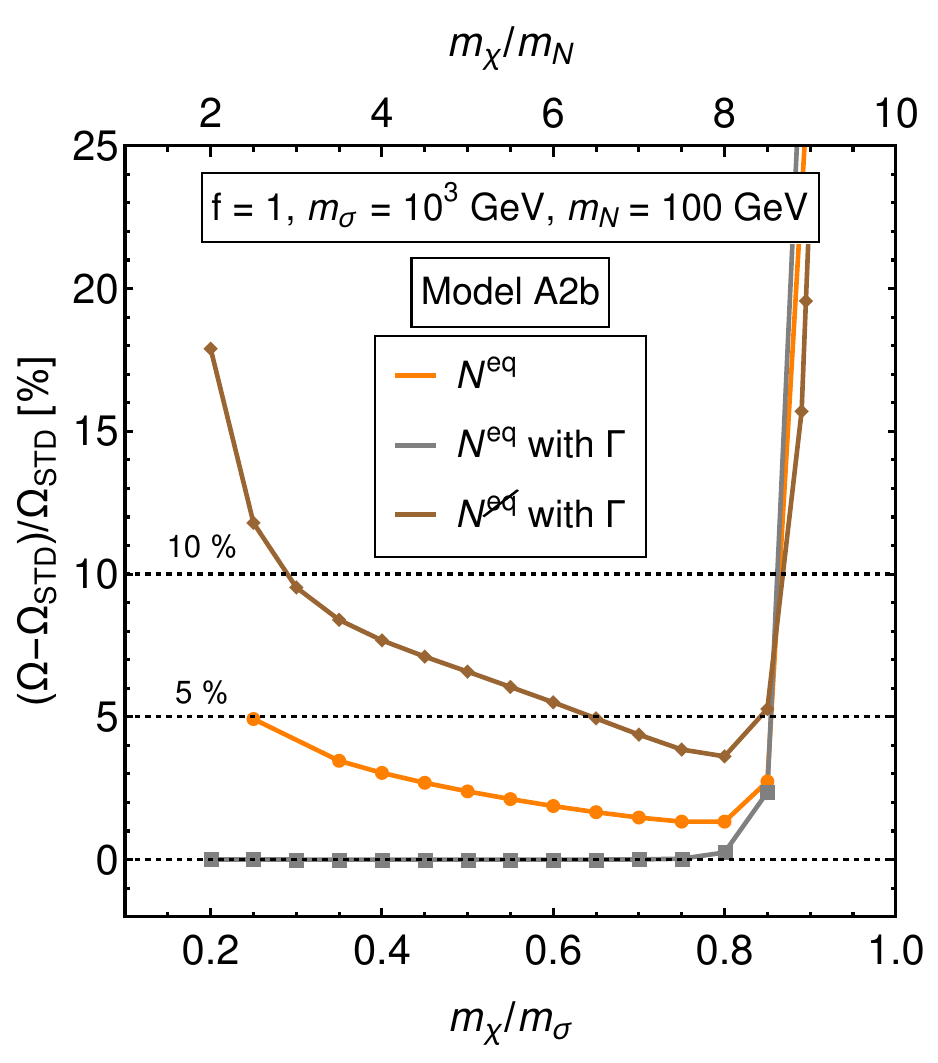}
 \caption{Measurement of the relative deviation in the relic abundance ($\Omega$) and the standard approximation ($\Omega_{\rm STD}$) calculated in different cases for Model A2b: solving the full set of BEqs in Eqs.~\eqref{eq:BEqs_1}--\eqref{eq:BEqs_4} (brown line), setting $N$ in equilibrium (orange line) and assuming also $N$ in equilibrium but taking into account decays (grey line). 
We have fixed $m_N=2\,\gev$ and $m_N=100\,\gev$ 
in the left and right panels, respectively.}
 \label{fig:ratios_omegas}
\end{figure}
In Fig.~\ref{fig:ratios_omegas}, 
the relative deviation of the BEqs' result from the STD approach is shown by plotting the quantity $(\Omega-\Omega_{\mathrm{STD}})/\Omega_{\mathrm{STD}}\,[\%]$ as a function of $m_\chi/m_{\sigma,N}$ for Model A2b, for given values of $f=1,\,m_\sigma=10^3\,\gev$ and 
$m_N = 2$~GeV (left) and 100~GeV (right), in three different cases: (i)~assuming that $N$ is in equilibrium, and following the evolution of $\sigma$ and $\chi$ (orange line); (ii)~same as before but taking into account $1 \leftrightarrow 2$ processes (grey line); and finally (iii)~the general case solving the BEqs for $\sigma,\,\chi$ and $N$, \textit{i.e.} assuming that $N$ is not in equilibrium, and also allowing for $1 \leftrightarrow 2$ processes (brown line). Notice that we show the case $m_N=100\,\gev$ in order to illustrate the case when masses for $\chi$ and $N$ are degenerate, $m_\chi \sim m_N$. Departures from $\Omega_{\mathrm{STD}}$ in the plot can be understood as follows.

From Eqs.~\eqref{eq:BEqs_1}--{\eqref{eq:BEqs_4}} it should be noted the inclusion of other particles, $\sigma$ and $N$, as evolving in temperature, so they can abandon the equilibrium like the DM. This feature was seen to be more significant when $N$ and/or $\sigma$ and the DM start to become non-relativistic at nearly the same temperature, so for similar masses. Then it tends to make the freeze-out happen earlier.\footnote{This was also noticed for the Model A1 in Refs.~\cite{Bandyopadhyay:2018qcv,Tang:2016sib}.} This is shown in the plot when the three lines depart from 
zero for $m_\chi \sim m_\sigma$ or when the brown line does so for $m_\chi \sim m_N$. 
Indeed, the latter feature is present only in the right panel of Fig.~\ref{fig:ratios_omegas}, where $m_N = 100$~GeV is relatively close to $m_\chi$.

Moreover, the addition of decay widths allows for the production/decay of $N$ from/to SM particles, and the decays of $\sigma$ to 
$\chi$ and $N$. This would produce, in contrast, the opposite effect by making the particles follow the equilibrium for longer if the decaying particle is not excessively Boltzmann suppressed. This can be noticed by the fact that the grey line is closer to zero than the orange line.
In conclusion, the deviation of the full BEqs' result from the STD 
is below 5\,\% (10\,\%) in almost all of the parameter space 
for $m_N = 2~(100)$~GeV, 
except for $m_\chi \sim m_\sigma$ ($m_\chi \sim m_{\sigma},\,m_N$).

\subsection{Departure from kinetic equilibrium of the dark sector with the SM}
%

\begin{figure}[t]
\centering
\includegraphics[width=.49\textwidth]{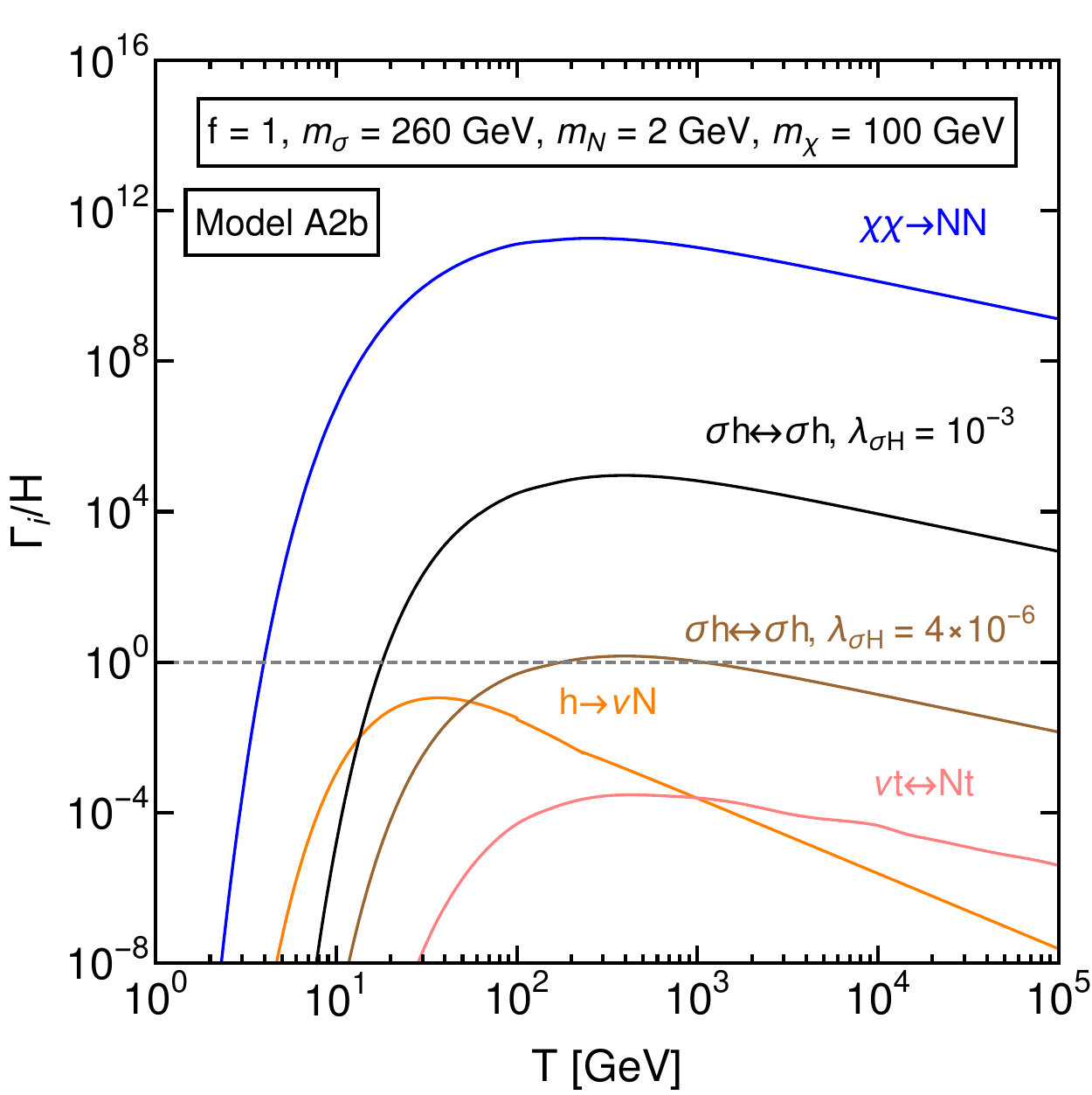}
\caption{Rates of the most relevant $1\leftrightarrow 2$ and $2 \leftrightarrow 2$ processes in Model A2b 
normalised to the Hubble rate.
The parameters have been fixed to $f=1$, $m_N=2$~GeV, $m_\chi=100$~GeV and $m_\sigma=260$~GeV. They correspond to a point in Fig.~\ref{fig:Results_plot}
yielding the correct value of relic abundance and avoiding all the experimental bounds.}
\label{fig:gamma_H}
\end{figure}

Here we briefly discuss kinetic decoupling of the dark sector 
from the SM. 
For concreteness, we focus on Model~A2b. 
In Fig.~\ref{fig:gamma_H}, we show the thermal rates of the most relevant 
$1 \leftrightarrow 2$ and $2 \leftrightarrow 2$ processes, normalised to the Hubble rate. The values of the fixed parameters correspond to a point 
in Fig.~\ref{fig:Results_plot} yielding the observed DM relic abundance
and avoiding all the experimental constraints.
As can be seen, the neutrino Yukawa coupling fixed by the seesaw relation 
in Eq.~\eqref{eq:seesaw_rel} to approximately $5 \times 10^{-8}$ 
(for $m_N = 2$~GeV and $m_\nu \simeq 0.05$~eV) cannot keep the dark sector in 
equilibrium with the SM. 
At the same time, the Higgs portal coupling $\lambda_{\sigma H}$
does ensure kinetic equilibrium between the dark sector and the SM as long as it is larger than $\sim 10^{-6}$, 
at least in some range of temperatures. 
For the example shown in Fig.~\ref{fig:gamma_H}, kinetic decoupling 
of the dark sector from the SM happens before the DM chemical freeze-out. However, kinetic equilibrium within the dark sector is maintained through $\chi N \leftrightarrow\chi N$ process.%
\footnote{Generally, this is the case for $m_N\lesssim m_\chi/20$, so that $N$ is relativistic at DM chemical freeze-out.}

From the moment of kinetic decoupling, the dark sector and the SM bath evolve with 
two different temperatures, $T_D$ and $T$, respectively. 
We assume that entropy is conserved independently in both sectors~\cite{Berlin:2016gtr}: 
\begin{equation}
\frac{s_D}{s_\mathrm{SM}} = \frac{s_D}{s_\mathrm{SM}} \bigg\rvert_{T=T_\mathrm{kd}}\,,
\end{equation}
with $s_D$ and $s_\mathrm{SM}$ being the entropy densities of the dark sector and the SM bath, respectively, and $T_\mathrm{kd}$ the temperature of kinetic decoupling. The evolution of $\xi \equiv T_{D}/T$ can be obtained using $s_\mathrm{SM} = (2\pi^2/45) g_{*}(T) T^3$ and $s_{D} = (\rho_D(T_D)+p_D(T_D))/T_D$, where $\rho_D$ and $p_D$ are the energy and pressure densities of the dark sector, and $g_{*}$ is the effective number of relativistic degrees of freedom in the visible sector.

\begin{figure}[t]
\centering
\includegraphics[width=.49\textwidth]{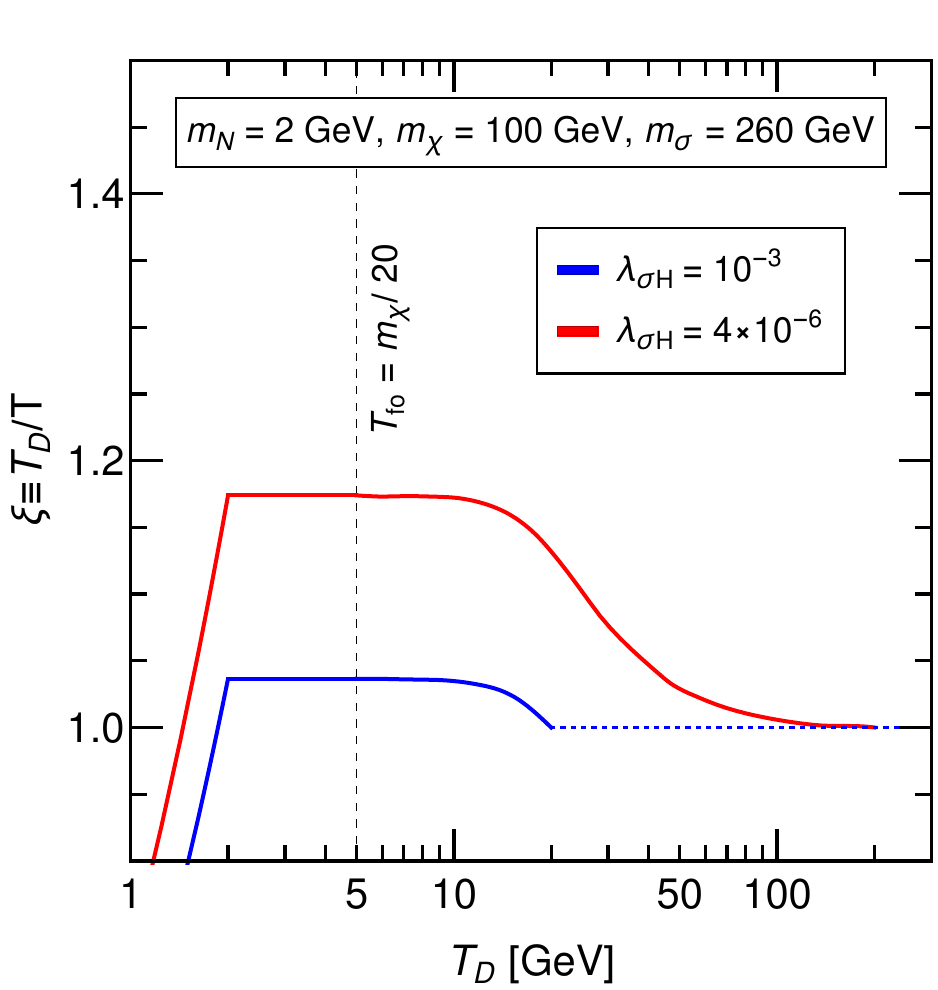}
\caption{Evolution of the ratio of the dark sector and SM temperatures, $\xi \equiv T_D/T$, 
for two representative values of the Higgs portal coupling $\lambda_{\sigma H}$ in Model A2b. 
The masses have been fixed to the values used in Fig.~\ref{fig:gamma_H}.
For $\lambda_{\sigma H} = 10^{-3}$ (blue line), $N$ is relativistic at the time of kinetic decoupling, $T_D \approx 20$~GeV, and down to the DM freeze-out at $T_D \approx 5$~GeV. 
For $\lambda_{\sigma H} = 4 \times 10^{-6}$ (red line), both $\chi$ and $N$ are relativistic 
at kinetic decoupling, $T_D \approx 200$~GeV, but only $N$ is at the freeze-out.}
\label{fig:xi_vs_Tdark}
\end{figure}

In Fig.~\ref{fig:xi_vs_Tdark}, we display the evolution of $\xi$ as function of the dark temperature, $T_D$, for the same point in the parameter space as in Fig.~\ref{fig:gamma_H}. We are interested in the value of $\xi$ at chemical freeze-out. 
For blue (red) line corresponding to the Higgs portal coupling $\lambda_{\sigma H}=10^{-3}~(4\times10^{-6})$, kinetic decoupling takes place at $T_D \approx 20~(200)$~GeV, cf.~also Fig.~\ref{fig:gamma_H}.
As can be seen from Fig.~\ref{fig:xi_vs_Tdark}, if $T_\mathrm{kd}<m_\chi$, the temperature of the dark sector
is very similar to that of the SM bath, whereas if $T_\mathrm{kd}>m_\chi$, the ratio of temperatures reaches approximately 1.2 at the freeze-out of DM.
In both cases, $N$ is relativistic at the freeze-out, 
and according to Ref.~\cite{Berlin:2016gtr}, the DM relic abundance 
is modified with respect to the standard solution by a factor 
$\xi\sqrt{g_*^\mathrm{eff}/g_*}$, 
where $g_*^\mathrm{eff} = g_* + g_D \xi^4$, with $g_D$ 
being the effective number of relativistic degrees of freedom in the dark sector. 
Since $g_D \ll g_*$ and $1 \lesssim \xi \lesssim 1.2$, 
the correction to the relic abundance 
can reach up to approximately 20\,\%.%
\footnote{For $m_\chi \lesssim 10$~GeV, $g_\ast$ 
decreases significantly due to the QCD transition, 
which has an additional impact on the correction.}
We have checked that this holds in a large part of the parameter space 
presented in Fig.~\ref{fig:Results_plot}; 
\textit{e.g.} if $\lambda_{\sigma H} = 10^{-3}$, 
this condition is fulfilled for $m_\chi \gtrsim 10$~GeV. 
After the freeze-out and until $N$ becomes non-relativistic, $\xi$ is constant, 
whereas after $T_D$ drops below $m_N$, the ratio $\xi \propto T$~\cite{Berlin:2016gtr}. 
More precisely, $\xi$ can be expressed as $\xi = \xi_\mathrm{fo} \sqrt{T_D/m_N}$, where $\xi_\mathrm{fo}$ 
is the value of $\xi$ at the freeze-out.

If sterile neutrinos become non-relativistic before the freeze-out, the dark sector may be significantly reheated. In that case, there could be an order one correction to the relic abundance. 
For a precise determination of the relic abundance in the presence of 
decoupled dark sectors and the cases where the impact of such a decoupling 
can be sizeable we refer the reader to Refs.~\cite{Binder:2017rgn,Bringmann:2020mgx}.

\bibliographystyle{JHEP}
\small
\setlength{\bibsep}{0pt}

\bibliography{nuRportals}

\end{document}